\documentclass[11pt]{article}

\usepackage[preprint]{acl}

\usepackage{times}
\usepackage{latexsym}

\usepackage[T1]{fontenc}
\usepackage[utf8]{inputenc}

\usepackage{microtype}

\usepackage{inconsolata}

\usepackage{graphicx}

\usepackage{graphicx}
\usepackage{xcolor}
\usepackage{amsmath}
\usepackage{pgfplots}
\usepgfplotslibrary{polar}
\pgfplotsset{compat=1.17}

\usepackage{booktabs}
\usepackage{pgfplots}
\usepgfplotslibrary{polar}

\usepackage{diagbox}

\usepackage[table,x11names,dvipsnames]{xcolor}
\usepackage{booktabs}
\usepackage{multirow}
\usepackage{amsmath}
\usepackage{amssymb}
\usepackage{minitoc}
\usepackage{titletoc}
\usepackage{enumitem}
\usepackage{subcaption}
\usepackage{threeparttable}

\usepackage{xcolor}
\usepackage{listings}
\usepackage{tcolorbox}
\tcbuselibrary{listings,skins,breakable}
\usepackage{cuted}

\usepackage{tikz}
\newcommand*\circled[1]{\tikz[baseline=(char.base)]{
            \node[shape=circle,draw,inner sep=0.4pt] (char) {#1};}}
            
\usepackage{tcolorbox}

\definecolor{nred}{RGB}{196, 38, 11}
\definecolor{ngreen}{RGB}{18, 141, 21}
\definecolor{nblue}{RGB}{41, 52, 190}
\definecolor{dartgreen}{HTML}{00693e}
\definecolor{uciblue}{HTML}{0064A4}
\definecolor{maroon}{cmyk}{0,0.87,0.68,0.32}

\newcommand{\thinktag}{\textcolor[HTML]{F38200}{\texttt{<think>}}}
\newcommand{\answertag}{\textcolor[HTML]{449F23}{\texttt{<answer>}}}

\hypersetup{
    colorlinks=true,
    linkcolor=uciblue,
    citecolor=uciblue,
    filecolor=magenta,      
    urlcolor=uciblue,
    }

\newtcblisting{promptbox}[1][]{
  enhanced,
  breakable,
  colback=uciblue!5,
  colframe=black,
  colbacktitle=uciblue!15,
  coltitle=black,
  fonttitle=\bfseries,
  title={#1},
  boxrule=0.6pt,
  arc=6pt,
  outer arc=6pt,
  left=6pt,right=6pt,top=6pt,bottom=6pt,
  listing only,
  listing options={
    basicstyle=\ttfamily\footnotesize\raggedright,
    breaklines=true,
    columns=fullflexible,
    keepspaces=true,
    showstringspaces=false
  }
}

\newcommand{\mname}{AudioLens-R1 }
\newcommand{\Mname}{AudioLens-R1}
\newcommand{\dname}{AudioLens-Bench }
\newcommand{\Dname}{AudioLens-Bench}

\title{AudioLens: Multi-Perspective Speech Clustering with Reasoning Audio-Language Models}

\author{
 \textbf{Wenjun Huang\textsuperscript{1}\thanks{Equal contribution.}},
 \textbf{Qiaosong Chu\textsuperscript{2}\footnotemark[1]},
 \textbf{Tiger Shao\textsuperscript{3}},
 \textbf{Pengfei Zhang\textsuperscript{1}},
\\
 \textbf{Yutong Song\textsuperscript{1}},
 \textbf{Hanning Chen\textsuperscript{1}},
 \textbf{Yezi Liu\textsuperscript{1}},
 \textbf{Weiyi Wu\textsuperscript{3}},
\\
 \textbf{SungHeon Jeong\textsuperscript{1}},
 \textbf{Ryozo Masukawa\textsuperscript{1}},
 \textbf{Sanggeon Yun\textsuperscript{1}},
 \textbf{Yang Ni\textsuperscript{4}},
\\
 \textbf{Jiang Gui\textsuperscript{3}},
 \textbf{Mohsen Imani\textsuperscript{1}},
\\
\\
 \textsuperscript{1}University of California, Irvine,
 \textsuperscript{2}Independent Researcher,\\
 \textsuperscript{3}Dartmouth College,
 \textsuperscript{4}Purdue University Northwest,
\\
}

\begin{document}
\maketitle
\begin{abstract} 

Audio clustering is a fundamental task for organizing rapidly growing speech collections, supporting applications such as conversational analysis and speech-driven discovery. 
However, existing methods rely on fixed acoustic similarity metrics or ASR-based text pipelines, limiting their ability to reorganize the same audio collection under different user-specified perspectives, especially when clustering depends on both linguistic and paralinguistic cues. 
We introduce \textbf{audio multi-perspective clustering}, where a model directly partitions speech recordings according to a natural-language perspective while inferring both the number of clusters and their assignments. 
To study this setting, we construct \Dname, a benchmark spanning multiple application domains and evaluating both in-perspective and cross-perspective generalization. 
We further propose \Mname, an end-to-end large audio-language model trained with reasoning distillation and preference optimization. 
Experiments show that \mname consistently outperforms all baselines, improving overall ARI by \textbf{12.99 points} and V-measure by \textbf{11.62 points}. 
These results demonstrate the promise of native audio-language models for flexible, perspective-conditioned structure discovery over speech collections.
\end{abstract}

% 我想着我写的话，可能我就先写audio clu是干啥的，然后有什么application，引几个不同的，然后我可能就直接讲，当前主要有两种方法来做这件事，teaser的，但是他们的limitation在哪，再讲用端到端audio llm做这个的优势是啥。

\section{Introduction}

Audio clustering aims to organize collections of speech recordings into coherent groups, and is a fundamental component for speech-driven analysis, retrieval, and discovery~\cite{park2022review,casanueva2020efficient,larson2012spoken,hu2023meetingbank,clifton2020podcasts}. 
As speech data rapidly grows in conversational agents, meetings, podcasts, and domain-specific audio archives, users increasingly need to reorganize the same collection according to different analytical goals. 
For example, the same set of recordings may be grouped by communicative intent, affective state, speaker-related traits, or background context. 
This requires clustering systems to flexibly adapt to user-specified criteria and to reason over both linguistic content and paralinguistic cues.

\begin{figure}[tb]
\centering
\resizebox{\columnwidth}{!}{
\includegraphics{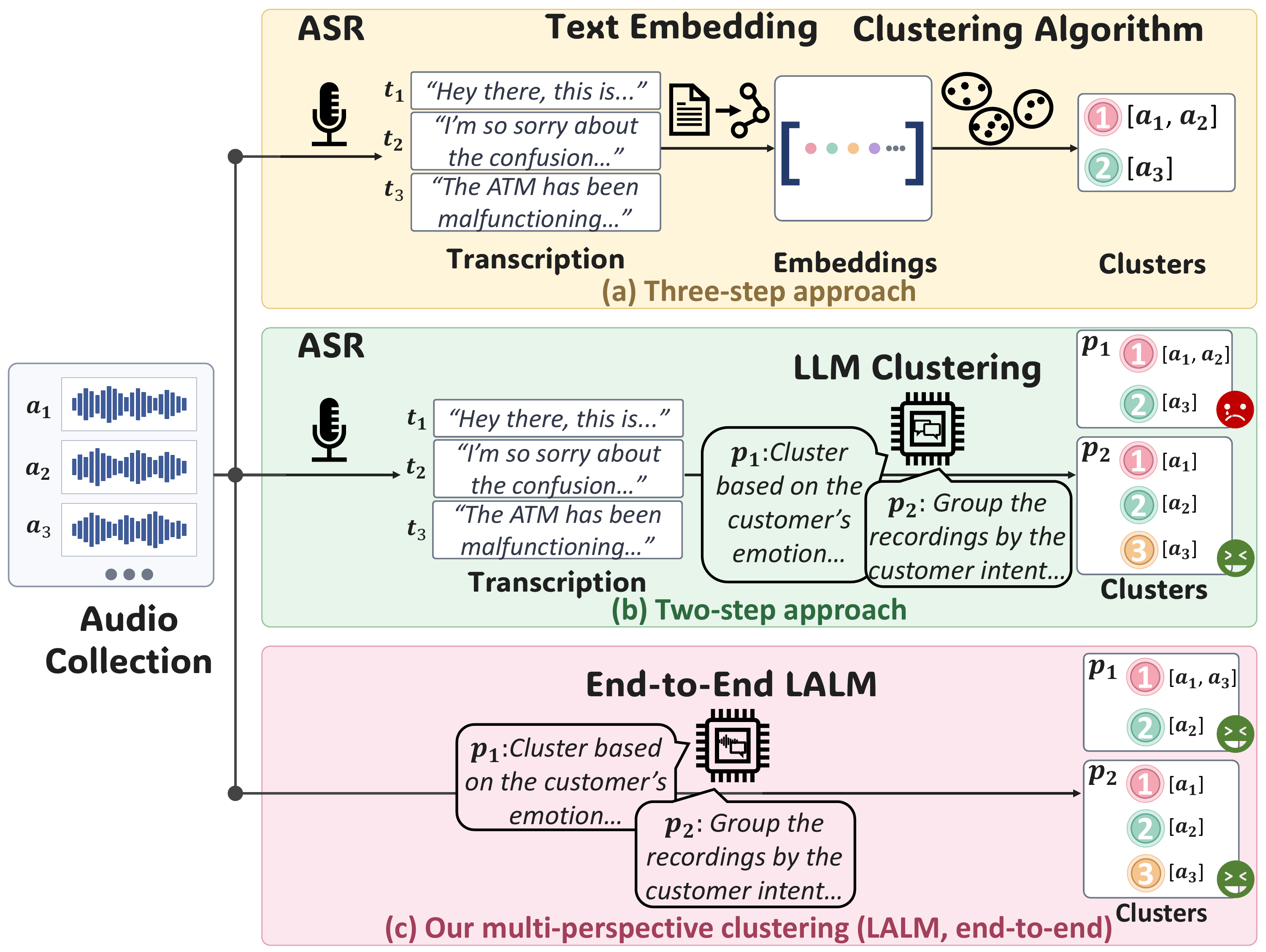}
}
\caption{
Three paradigms for audio clustering.
(a) Traditional pipelines transcribe audio, embed transcripts, and apply a clustering algorithm.
(b) Transcript-based LLM clustering follows user-specified perspectives, but remains limited when the perspective depends on paralinguistic cues.
(c) \mname takes raw audio and natural-language perspectives as input, enabling the same speech collection to be clustered into different valid partitions under different perspectives.
}
\label{fig:teaser}
\vspace{-1em}
\end{figure}

Existing approaches are limited in this setting. 
Traditional acoustic clustering methods operate directly on speech representations and are effective for predefined criteria such as speaker identity or acoustic similarity, but they typically rely on fixed similarity metrics and task-specific representations. 
Transcript-centric pipelines, which are the focus of the comparison in Fig.~\ref{fig:teaser}(a), provide a more semantic alternative.
They transcribe each recording with an automatic speech recognition (ASR) model, encode the transcript with a sentence encoder~\cite{reimers2019sentence}, and apply a classical clustering algorithm such as K-Means~\cite{lloyd1982least} or Gaussian mixture models (GMMs)~\cite{dempster1977maximum}. 
A more recent line of work (Fig.~\ref{fig:teaser}(b)) replaces the clustering stage with a large language model (LLM) that performs clustering directly over the transcripts~\cite{zhang2023clusterllm,viswanathan2024large}. 
Yet transcript-centric pipelines share a fundamental bottleneck: \textbf{ASR captures \textit{what} is said but discards much of \textit{how} it is delivered}, including prosody, speaker traits, etc. 
As a result, existing methods either specialize in fixed acoustic notions of similarity or reason over text-only representations, but lack a unified mechanism for reorganizing the same speech collection under flexible perspectives that depend on both linguistic and paralinguistic information.

Recent large audio-language models (LALMs) provide a promising foundation for this problem because they can process speech directly and jointly model semantic and acoustic information~\cite{diao2025soundmind}. 
Nevertheless, current LALMs are primarily evaluated on audio understanding and generation, rather than on structure discovery over speech collections. 
It remains unclear whether such models can interpret a natural-language clustering perspective, compare multiple audio segments under that perspective, infer the appropriate number of clusters, and produce a valid partition without relying on a separate clustering algorithm.

In this work, we introduce \textbf{audio multi-perspective clustering}, a new task that formulates audio clustering as perspective-conditioned structure discovery. 
Given a set of speech recordings and a natural-language clustering perspective, the model must infer both the number of clusters and the assignment of each recording. 
Unlike conventional settings where the similarity function or the number of clusters is fixed in advance, our setting allows the same audio collection to induce different valid partitions under different perspectives. 
This formulation directly tests whether models can use natural-language criteria to select the relevant linguistic and paralinguistic evidence for clustering.

To support this task, we construct \Dname, \textbf{the first benchmark for multi-perspective audio clustering across diverse application domains.}
It includes perspectives grounded in lexical-semantic reasoning as well as paralinguistic attributes, enabling evaluation of both text-like reasoning and audio-native perception. 
We further organize the benchmark into held-in and held-out perspectives, allowing us to measure in-perspective generalization under familiar criteria and cross-perspective generalization to unseen criteria.

We also propose \textbf{\Mname, an end-to-end LALM-based clustering model }depicted in Fig.~\ref{fig:teaser}(c). 
\Mname{} takes raw audio segments and a natural-language perspective as input, and directly generates a structured clustering answer. 
To adapt LALMs to this task, we first train the model with reasoning distillation, which provides comparison-based supervision for perspective-conditioned clustering. 
We then apply direct preference optimization using valid but incorrect model-generated partitions as hard negatives, encouraging the model to prefer clustering decisions that better match the gold partition while preserving output validity.

Our contributions are summarized as follows:
\begin{itemize}[leftmargin=*]
\vspace{-.7em}
    \item We formalize \textbf{audio multi-perspective clustering}, where speech recordings are clustered according to a natural-language perspective and the model must infer both cluster number and assignments.
    \vspace{-.7em}
    \item We introduce \Dname, a benchmark covering diverse domains, linguistic and paralinguistic perspectives, and held-in / held-out perspective splits.
    \vspace{-.7em}
    \item We propose \Mname, an end-to-end LALM trained with reasoning distillation and preference optimization for perspective-conditioned audio clustering.
    \vspace{-.7em}
    \item Experiments demonstrate that \Mname{} consistently outperforms the baselines, improving overall ARI by \textbf{12.99} points and V-measure by \textbf{11.62} points.
\end{itemize}

\section{Benchmark Construction}
\label{sec:benchmark}

We introduce \Dname, a benchmark for evaluating whether models can cluster speech recordings according to natural-language perspectives. 
The benchmark is designed around two requirements. 
First, the clustering criterion should be flexible: the same audio collection may induce different valid partitions under different perspectives. 
Second, the benchmark should require both linguistic and paralinguistic reasoning, since many realistic audio-organization scenarios depend not only on what is said but also on how it is spoken.

\begin{figure}[tb]
\centering
\resizebox{\columnwidth}{!}{
\includegraphics{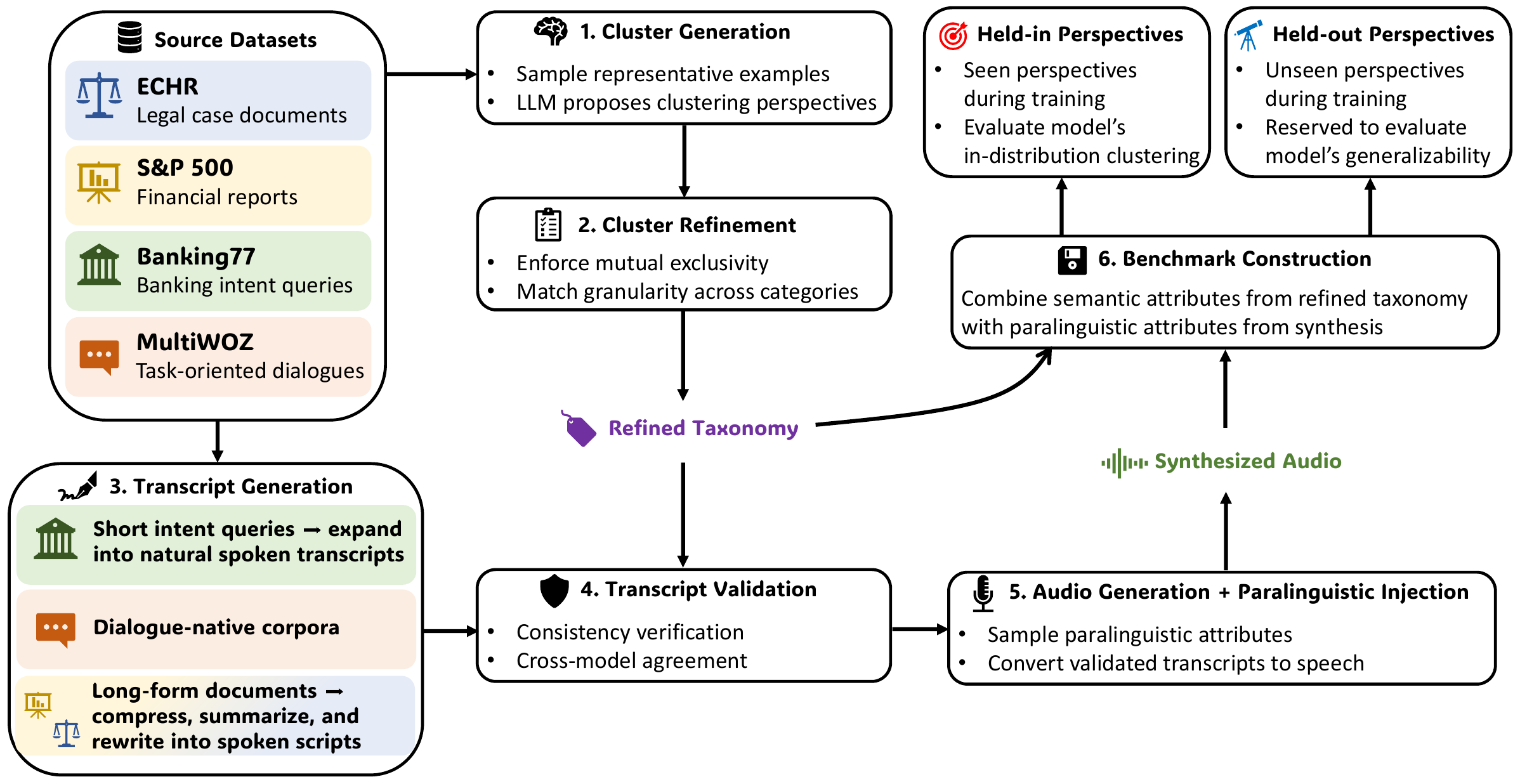}
}
\caption{
Overview of \dname construction.
}
\label{fig:benchmark_pipeline}
\vspace{-1em}
\end{figure}

\subsection{Source Corpora}

We construct \Dname{} from four complementary corpora: ECHR~\citep{poudyal2020echr}, S\&P 500 annual reports~\citep{jlohding_sp500_edgar_10k}, Banking77~\citep{casanueva2020efficient}, and MultiWOZ~\citep{zang2020multiwoz}. 
These corpora cover legal cases, financial disclosures, banking service requests, and task-oriented dialogues, providing diverse domains and reasoning patterns. 
The original corpora are text-based or dialogue-based, so we convert them into speech recordings while preserving the clustering evidence required by each perspective. 
Corpus-specific construction details and statistics are provided in App.~\ref{app:corpus_pipeline}.

\subsection{Perspective and Audio Synthesis}

Fig.~\ref{fig:benchmark_pipeline} illustrates the construction pipeline. 
For each corpus, we first induce candidate clustering perspectives from representative examples using an LLM. 
Each perspective defines a clustering criterion and a set of mutually exclusive categories. 
We then refine the proposed perspectives to ensure that categories are interpretable, comparable in granularity, and sufficiently supported by examples. 
After refinement, each retained perspective is converted into a natural-language clustering instruction that does not reveal category names.

We next construct speech instances aligned with these perspectives. 
Depending on the source corpus, we either expand short user queries into spoken scripts, compress long documents into evidence-preserving spoken summaries, or reuse dialogue-native transcripts. 
To reduce lexical shortcuts, generated transcripts are not allowed to explicitly mention perspective names, category names, or near-verbatim instruction phrases. 
We validate each transcript by re-classifying it under the corresponding perspective and retaining only instances whose labels are consistently recovered.

Finally, validated transcripts are synthesized into speech. 
During synthesis, we inject controlled paralinguistic attributes such as emotion, speaker identity, speaker count, and background acoustic condition. 
For linguistic perspectives, these attributes are randomized and approximately balanced across categories to avoid spurious correlations. 
For paralinguistic perspectives, the target acoustic attribute defines the clustering criterion, while other attributes are randomized. 
This design allows \Dname{} to evaluate both lexical-semantic clustering and audio-native clustering.

To minimize potential bias during dataset construction, three co-authors independently reviewed the samples in \Dname. 
Specifically, they assessed the relevance of the clustering perspective, transcript accuracy, audio intelligibility, semantic-label correctness, and perceptibility of the injected paralinguistic attributes. 
A sample was retained only when the co-authors agreed that it satisfied these criteria; otherwise, it was discarded and regenerated.
\vspace{-.5em}
\subsection{Benchmark Organization}

We split clustering perspectives into \textit{held-in} and \textit{held-out} perspectives. 
Held-in perspectives are observed during training, including their instructions and underlying taxonomies. 
Held-out perspectives are reserved exclusively for evaluation, and their instructions, categories, and instances are never used during training. 
This perspective-level split allows us to distinguish in-perspective generalization from cross-perspective generalization.

For held-in perspectives, audio recordings within each category are divided into a training pool and an evaluation pool. 
Training instances are created by sampling categories and then sampling audio recordings from each selected category. 
Evaluation instances are organized into three levels: 
\circled{1} $\mathrm{\textbf{L}}_0$: seen perspectives and seen audio recordings, but unseen category/audio combinations. This evaluates recombination robustness.
\circled{2} $\mathrm{\textbf{L}}_1$: seen perspectives but unseen audio recordings and unseen combinations. This evaluates generalization to new audio under familiar perspectives.
\circled{3} $\mathrm{\textbf{L}}_2$: unseen perspectives, unseen audio recordings, and unseen combinations. This evaluates cross-perspective generalization.
The statistics of \Dname{} are summarized in Tab.~\ref{tab:dataset_statistics}.
Details about the benchmark organization are provided in App.~\ref{app:split_construction}.
\begin{table}[t]
\centering
\small
\setlength{\tabcolsep}{5.0pt}
\renewcommand{\arraystretch}{1.08}
\resizebox{\columnwidth}{!}{
\begin{tabular}{lcccccccc}
\toprule
\toprule
\multirow{2}{*}{Dataset} 
& \multicolumn{2}{c}{Train}
& \multicolumn{2}{c}{$\mathrm{\textbf{L}}_0$}
& \multicolumn{2}{c}{$\mathrm{\textbf{L}}_1$}
& \multicolumn{2}{c}{$\mathrm{\textbf{L}}_2$} \\
\cmidrule(lr){2-3}
\cmidrule(lr){4-5}
\cmidrule(lr){6-7}
\cmidrule(lr){8-9}
& \#Audio & \#Data
& \#Audio & \#Data
& \#Audio & \#Data
& \#Audio & \#Data \\
\midrule
\rowcolor{uciblue!5}
Banking77
& 641 & 4110
& 584 & 1863
& 519 & 1800
& 599 & 1849 \\

MultiWOZ
& 466 & 1571
& 479 & 1617
& 440 & 1574
& 499 & 1578 \\
\rowcolor{uciblue!5}
ECHR
& 519 & 3795
& 478 & 1735
& 453 & 1665
& 478 & 1670 \\

S\&P 500
& 472 & 3907
& 404 & 1784
& 373 & 1700
& 396 & 1689 \\

\bottomrule
\bottomrule
\end{tabular}
}
\caption{
Statistics of \Dname. \#Audio denotes the number of unique audio recordings, and \#Data denotes the number of clustering instances.
}
\vspace{-2em}
\label{tab:dataset_statistics}
\end{table}

\vspace{-.5em}
\section{Method}
\label{sec:method}
% \vspace{-1em}
We propose \Mname, an end-to-end large audio-language model for audio multi-perspective clustering. 
The model is trained in two stages: reasoning distillation, which teaches comparison-based clustering behavior, and preference optimization, which further aligns the model toward higher-quality clustering decisions.

\subsection{Problem Formulation}
\label{sec:formulation}

Let $\mathcal{D}=\{a_1,a_2,\ldots,a_n\}$ denote a collection of speech segments, where each $a_i$ is an audio recording. 
Let $p$ denote a natural-language clustering perspective that specifies the criterion for grouping the recordings. 
The goal is to produce a partition
\begin{equation}
\mathcal{C}=\{C_1,C_2,\ldots,C_K\},
\end{equation}
where each $C_k \subseteq [n]$ is a non-empty cluster and the partition satisfies
\begin{equation}
C_k \cap C_{k'}=\emptyset \quad (k\neq k'), 
\qquad
\bigcup_{k=1}^{K} C_k = [n].
\end{equation}
The number of clusters $K$ is not provided as input and must be inferred by the model:
\begin{equation}
(K,\mathcal{C}) = f_\theta(\mathcal{D},p).
\end{equation}

Because clustering is permutation-invariant, cluster names and cluster order do not affect correctness. 
We therefore evaluate model outputs as unordered partitions over item indices. 
In practice, each audio segment is serialized with a 1-based index, and the model is required to generate a valid answer block that assigns every index exactly once.

\subsection{Reasoning Distillation}
\label{sec:rd}

Directly training an LALM to output final partitions can encourage shallow pattern imitation and does not explicitly teach the model how to compare audio segments under a perspective. 
We therefore first perform reasoning distillation (RD). 
For each training instance, we construct a gold partition from the benchmark annotations and ask a teacher model to synthesize a concise reasoning trace that leads to the gold clustering decision.
The resulting distillation dataset is
\begin{equation}
\mathcal{D}_{\mathrm{RD}}
=
\{(x^{(i)}, y_{\mathrm{trace}}^{(i)})\}_{i=1}^{M},
\end{equation}
where $x^{(i)}=(\mathcal{D}^{(i)},p^{(i)})$ contains the indexed audio set and the clustering perspective, and $y_{\mathrm{trace}}^{(i)}$ contains a reasoning trace followed by the final clustering answer. 
We generate complementary traces for linguistic and paralinguistic perspectives: linguistic traces are grounded in transcribed content, while paralinguistic traces are grounded in audible cues. 
All traces are filtered to ensure that their final partitions match the gold partition and that the reasoning is comparison-based, concise, and modality-consistent. 
More details are provided in App.~\ref{app:rd_details}.

We train the model with standard autoregressive supervised fine-tuning:
\begin{equation}
\mathcal{L}_{\mathrm{RD}}(\theta)
=
-
\sum_{(x,y)\in \mathcal{D}_{\mathrm{RD}}}
\sum_{t=1}^{|y|}
\log p_\theta(y_t \mid x,y_{<t}).
\end{equation}
This stage teaches the model to interpret the perspective, compare recordings, infer the number of clusters, and produce a structurally valid partition.

\label{sec:reasoning_distillation}
\begin{figure}[tb]
\centering
\resizebox{\columnwidth}{!}{
\includegraphics{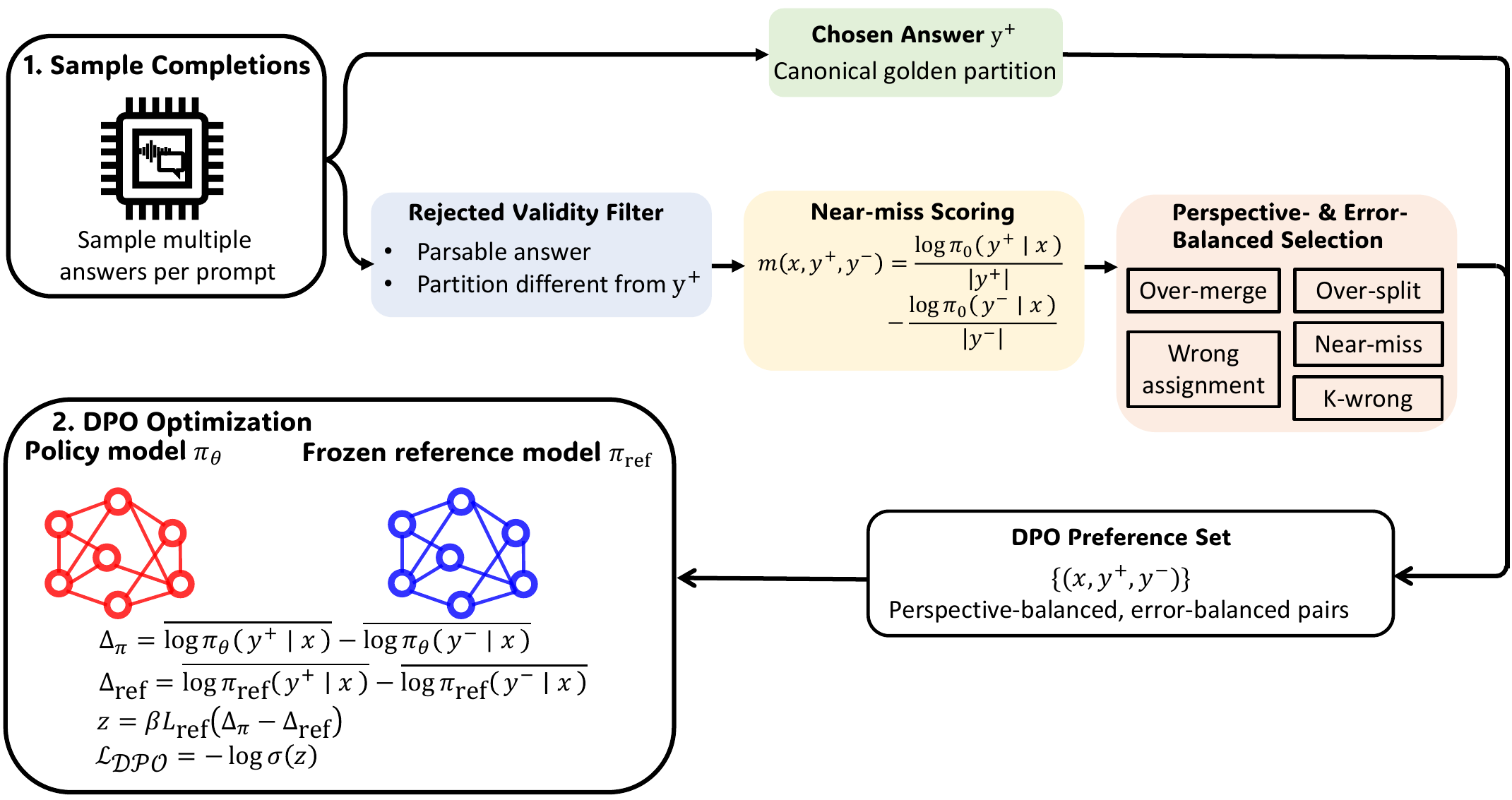}
}
\caption{
Overview of preference optimization pipeline.
}
\label{fig:dpo_pipeline}
\vspace{-0.5cm}
\end{figure}

\subsection{Preference Optimization}
\label{sec:dpo}

After reasoning distillation, we further optimize the model with Direct Preference Optimization (DPO) as illustrated in Fig.~\ref{fig:dpo_pipeline}. 
The goal is to improve clustering decisions while preserving the output format. 
For each input $x$, we construct a preference pair $(x,y^+,y^-)$, where $y^+$ is the canonical gold partition and $y^-$ is a valid but incorrect partition sampled from the RD model. 
We only keep rejected answers that are parsable, assign every item exactly once, and differ from the gold partition. 
This focuses preference learning on clustering errors rather than formatting failures.

To select informative pairs, we prioritize hard negatives that are close to the gold answer under the initial model or represent common structural clustering errors, such as over-merging, over-splitting, wrong cluster counts, or wrong item assignments. 
We also balance the selected pairs across perspectives and error types to avoid overfitting to a narrow class of mistakes. 
The detailed hard-pair selection procedure is described in App.~\ref{app:dpo_details}.

\begin{table*}[t!]
\centering
\resizebox{\linewidth}{!}{
\begin{tabular}{llcccccccccccccccc}
\toprule\toprule
\multirow{2}{*}{Method} &
\multirow{2}{*}{Clustering} &
  \multicolumn{3}{c}{ECHR} &
  \multicolumn{3}{c}{S\&P 500} &
  \multicolumn{3}{c}{Banking77} &
  \multicolumn{3}{c}{MultiWOZ} &
  \multicolumn{3}{c}{Average} &
  \multicolumn{1}{c}{Overall} \\ 
  \cmidrule(lr){3-5}\cmidrule(lr){6-8}\cmidrule(lr){9-11}\cmidrule(lr){12-14}\cmidrule(lr){15-17}
 &
 &
  \multicolumn{1}{c}{$\mathrm{\textbf{L}}_0$} &
  \multicolumn{1}{c}{$\mathrm{\textbf{L}}_1$} &
  \multicolumn{1}{c}{$\mathrm{\textbf{L}}_2$} &
  \multicolumn{1}{c}{$\mathrm{\textbf{L}}_0$} &
  \multicolumn{1}{c}{$\mathrm{\textbf{L}}_1$} &
  \multicolumn{1}{c}{$\mathrm{\textbf{L}}_2$} &
  \multicolumn{1}{c}{$\mathrm{\textbf{L}}_0$} &
  \multicolumn{1}{c}{$\mathrm{\textbf{L}}_1$} &
  \multicolumn{1}{c}{$\mathrm{\textbf{L}}_2$} &
  \multicolumn{1}{c}{$\mathrm{\textbf{L}}_0$} &
  \multicolumn{1}{c}{$\mathrm{\textbf{L}}_1$} &
  \multicolumn{1}{c}{$\mathrm{\textbf{L}}_2$} &
  \multicolumn{1}{c}{$\mathrm{\textbf{L}}_0$} &
  \multicolumn{1}{c}{$\mathrm{\textbf{L}}_1$} &
  \multicolumn{1}{c}{$\mathrm{\textbf{L}}_2$} &
   \\ \midrule
\multicolumn{18}{l}{\textbf{\textit{ASR (Whisper~\cite{radford2023robust}) + Text Embedding + Clustering}}} \\ \midrule
\rowcolor{uciblue!5}
& K-Means & 4.98 & 4.24 & 8.82 & 9.90 & 5.71 & 5.42 & 8.29 & 6.83 & 6.83 & 4.73 & 4.77 & 8.81 & 6.97 & 5.39 & 7.47 & 6.61 \\
\rowcolor{uciblue!5}
\multirow{-2}{*}{\cellcolor{uciblue!5}InBedder~\cite{peng2024answer}}
& GMM & 1.58 & 2.56 & 9.06 & 2.89 & 2.15 & 3.73 & 5.20 & 8.11 & 9.57 & 5.43 & 3.92 & 7.16 & 3.77 & 4.18 & 7.38 & 5.11 \\

& K-Means & 12.18 & 15.40 & 26.51 & 16.19 & 18.79 & 15.56 & 23.57 & 24.68 & 18.18 & 10.55 & 11.41 & 11.09 & 15.62 & 17.57 & 17.84 & 17.01 \\
\multirow{-2}{*}{Instructor~\cite{su2023one}}
& GMM & 8.44 & 9.68 & 17.30 & 9.37 & 9.46 & 6.98 & 14.79 & 13.13 & 14.65 & 7.30 & 7.66 & 6.48 & 9.97 & 9.98 & 11.35 & 10.44 \\

\rowcolor{uciblue!5}
& K-Means & 11.70 & 9.89 & 15.37 & 11.20 & 11.92 & 7.26 & 16.94 & 16.62 & 16.90 & 8.55 & 12.66 & 8.34 & 12.10 & 12.77 & 11.97 & 12.28 \\
\rowcolor{uciblue!5}
\multirow{-2}{*}{\cellcolor{uciblue!5}Qwen3-Embedding~\cite{zhang2025qwen3}}
& GMM & 9.53 & 6.97 & 13.68 & 8.63 & 5.73 & 3.77 & 8.38 & 11.39 & 12.36 & 3.96 & 5.88 & 5.07 & 7.62 & 7.49 & 8.72 & 7.95 \\

& K-Means & 12.93 & 14.32 & 24.95 & 16.19 & 14.36 & 13.89 & 19.72 & 23.28 & 18.51 & 9.81 & 9.96 & 14.47 & 14.66 & 15.48 & 17.96 & 16.03 \\
\multirow{-2}{*}{all-MiniLM-L6-v2~\cite{wang2020minilm}}
& GMM & 4.79 & 7.63 & 15.99 & 7.57 & 6.81 & 8.01 & 10.52 & 12.76 & 15.67 & 6.84 & 7.08 & 8.10 & 7.43 & 8.57 & 11.94 & 9.31 \\ \midrule
\multicolumn{18}{l}{\textbf{\textit{ASR (Whisper~\cite{radford2023robust}) + LLM}}} \\ \midrule
\rowcolor{uciblue!5}
GPT-4o~\cite{hurst2024gpt}
& -- &14.80&20.26&32.75&26.19&\underline{27.78}&26.30&\underline{32.53}&31.05&32.04&36.32&35.50&24.99&27.46&28.65&29.02&28.38\\
\midrule
\multicolumn{18}{l}{\textbf{\textit{LALM}}} \\ \midrule

\rowcolor{uciblue!5}
GPT-4o-audio-preview
& -- &\underline{19.34}&\underline{21.82}&38.26&\underline{28.01}&26.93&24.57&32.32&\underline{32.12}&40.42&34.90&34.90&28.27&\underline{28.64}&\underline{28.94}&32.88&30.16\\

GPT-audio-1.5
& -- &18.76&20.20&\underline{40.24}&25.56&27.75&\underline{26.71}&29.77&29.57&\underline{46.79}&\underline{37.99}&\underline{37.21}&\underline{40.82}&28.02&28.68&\underline{38.64}&\underline{31.78}\\

\rowcolor{uciblue!5}
Qwen3-omni-instruct-30B
& -- & 7.38 & 8.72 & 11.32 & 14.57 & 16.13 & 13.07 & 26.27 & 28.40 & 20.54 & 14.30 & 18.54 & 10.53 & 15.63 & 17.95 & 13.87 & 15.81 \\

Audio Flamingo 3
& -- & 0.00 & 0.00 & 0.00 & 0.00 & 0.00 & 0.00 & 0.00 & 0.00 & 0.00 & 0.57 & 1.17 & 0.98 & 0.14 & 0.29 & 0.25 & 0.23 \\

\rowcolor{uciblue!5}
Qwen2.5-omni
& -- & 6.30 & 8.37 & 5.59 & 6.31 & 6.36 & 4.87 & 6.37 & 7.68 & -0.79 & 13.97 & 15.25 & 8.58 & 8.24 & 9.42 & 4.56 & 7.41 \\

\rowcolor{uciblue!15}
\Mname
& -- &\textcolor{blue}{\textbf{41.91}}&\textcolor{blue}{\textbf{38.05}}&\textcolor{blue}{\textbf{46.11}}&\textcolor{blue}{\textbf{42.20}}&\textcolor{blue}{\textbf{44.86}}&\textcolor{blue}{\textbf{41.12}}&\textcolor{blue}{\textbf{52.23}}&\textcolor{blue}{\textbf{50.24}}&\textcolor{blue}{\textbf{47.07}}&\textcolor{blue}{\textbf{47.28}}&\textcolor{blue}{\textbf{44.14}}&\textcolor{blue}{\textbf{42.08}}&\textcolor{blue}{\textbf{45.91}}&\textcolor{blue}{\textbf{44.32}}&\textcolor{blue}{\textbf{44.10}}&\textcolor{blue}{\textbf{44.77}}\\

\bottomrule\bottomrule
\end{tabular}
}
\caption{ARI comparison across four corpora.
\mname achieves the best overall ARI and demonstrates strong performance across most clustering settings.
For embedding-based baselines, we report results with both K-Means and GMM clustering, while LLM/LALM-based methods directly produce clustering assignments.
The best result is highlighted in \textbf{\textcolor{blue}{bold blue}}, and the second-best result is \underline{underlined}.}
\vspace{-1em}
\label{tab:ARI}
\end{table*}

For each preference pair, we compute answer-token mean log-probabilities under the policy model $\pi_\theta$ and the frozen reference model $\pi_{\mathrm{ref}}$. 
Let
\begin{align}
&\Delta_\pi
=
\log \pi_\theta(y^+\mid x)
-
\log \pi_\theta(y^-\mid x),\notag \\
&\Delta_{\mathrm{ref}}
=
\log \pi_{\mathrm{ref}}(y^+\mid x)
-
\log \pi_{\mathrm{ref}}(y^-\mid x).
\end{align}
The DPO logit is
\begin{equation}
z = \beta L_{\mathrm{ref}}(\Delta_\pi-\Delta_{\mathrm{ref}}),
\end{equation}
where $L_{\mathrm{ref}}$ is the average answer length in the DPO training set. 
The preference loss is
\begin{equation}
\mathcal{L}_{\mathrm{DPO}}
=
-\log \sigma(z).
\end{equation}
All log-probabilities are computed only over the canonical clustering answer. 
Thus, reasoning traces provide intermediate supervision during RD, while DPO directly optimizes the final decision.

\section{Experimental Results}
\begin{table*}[]
\centering
\resizebox{\linewidth}{!}{
\begin{tabular}{llcccccccccccccccc}
\toprule\toprule
\multirow{2}{*}{Method} &
\multirow{2}{*}{Clustering} &
  \multicolumn{3}{c}{ECHR} &
  \multicolumn{3}{c}{S\&P 500} &
  \multicolumn{3}{c}{Banking77} &
  \multicolumn{3}{c}{MultiWOZ} &
  \multicolumn{3}{c}{Average} &
  \multicolumn{1}{c}{Overall} \\ 
  \cmidrule(lr){3-5}\cmidrule(lr){6-8}\cmidrule(lr){9-11}\cmidrule(lr){12-14}\cmidrule(lr){15-17}
 &
 &
\multicolumn{1}{c}{$\mathrm{\textbf{L}}_0$} &
  \multicolumn{1}{c}{$\mathrm{\textbf{L}}_1$} &
  \multicolumn{1}{c}{$\mathrm{\textbf{L}}_2$} &
  \multicolumn{1}{c}{$\mathrm{\textbf{L}}_0$} &
  \multicolumn{1}{c}{$\mathrm{\textbf{L}}_1$} &
  \multicolumn{1}{c}{$\mathrm{\textbf{L}}_2$} &
  \multicolumn{1}{c}{$\mathrm{\textbf{L}}_0$} &
  \multicolumn{1}{c}{$\mathrm{\textbf{L}}_1$} &
  \multicolumn{1}{c}{$\mathrm{\textbf{L}}_2$} &
  \multicolumn{1}{c}{$\mathrm{\textbf{L}}_0$} &
  \multicolumn{1}{c}{$\mathrm{\textbf{L}}_1$} &
  \multicolumn{1}{c}{$\mathrm{\textbf{L}}_2$} &
  \multicolumn{1}{c}{$\mathrm{\textbf{L}}_0$} &
  \multicolumn{1}{c}{$\mathrm{\textbf{L}}_1$} &
  \multicolumn{1}{c}{$\mathrm{\textbf{L}}_2$} &
   \\ 
\midrule
\multicolumn{18}{l}{\textbf{\textit{ASR (Whisper~\cite{radford2023robust}) + Text Embedding + Clustering}}} \\ \midrule
\rowcolor{uciblue!5}
& K-Means &50.71&51.51&57.84&53.69&50.88&48.79&52.13&52.87&50.09&30.24&29.38&29.55&46.69&46.16&46.57&46.47 \\

\rowcolor{uciblue!5}
\multirow{-2}{*}{\cellcolor{uciblue!5}InBedder~\cite{peng2024answer}}
& GMM &49.19&50.79&58.20&50.64&49.54&48.81&50.71&54.13&53.02&31.49&29.43&28.97&45.51&45.97&47.25&46.24 \\

\multirow{2}{*}{Instructor~\cite{su2023one}}
& K-Means &55.50&57.76&67.36&57.81&\underline{59.13}&54.67&60.18&61.91&57.21&34.75&34.41&31.18&52.06&53.30&52.61&52.66 \\
& GMM &53.50&55.07&62.85&54.41&53.58&50.40&56.26&57.21&55.96&32.98&32.15&28.51&49.29&49.50&49.43&49.41 \\

\rowcolor{uciblue!5}
& K-Means &55.12&54.87&61.72&55.11&54.51&50.05&57.61&58.61&56.49&33.51&35.38&29.19&50.34&50.84&49.36&50.18 \\
\rowcolor{uciblue!5}
\multirow{-2}{*}{\cellcolor{uciblue!5}Qwen3-Embedding~\cite{zhang2025qwen3}}
& GMM &54.34&53.33&61.20&54.18&51.45&48.67&53.03&56.43&54.61&30.92&30.75&27.60&48.12&47.99&48.02&48.04 \\

\multirow{2}{*}{all-MiniLM-L6-v2~\cite{wang2020minilm}}
& K-Means &\underline{55.68}&57.33&66.83&\underline{57.97}&56.61&53.83&58.34&61.16&57.07&34.22&33.67&33.55&51.55&52.19&52.82&52.19 \\
& GMM &51.43&53.67&62.37&53.29&52.22&51.26&53.81&57.34&56.69&32.83&31.83&30.10&47.84&48.77&50.10&48.90 \\

\midrule\multicolumn{18}{l}{\textbf{\textit{ASR (Whisper~\cite{radford2023robust}) + LLM}}} \\ \midrule
\rowcolor{uciblue!5}
GPT-4o~\cite{hurst2024gpt}
& -- &55.41&\underline{59.70}&69.98&56.03&56.80&57.39&\underline{67.81}&\underline{66.29}&70.85&\underline{57.75}&55.59&51.47&\underline{59.25}&\underline{59.60}&62.42&60.42 \\ \midrule
\multicolumn{18}{l}{\textbf{\textit{LALM}}} \\ \midrule

\rowcolor{uciblue!5}
GPT-4o-audio-preview~\cite{openai_gpt4o_audio_preview}
& -- &55.13&58.51&72.79&56.47&57.96&60.86&65.65&65.94&\underline{78.73}&55.53&54.16&53.94&58.19&59.14&66.58&61.30\\

GPT-audio-1.5~\cite{openai_gpt_audio_1_5}
& -- &53.73&53.60&\underline{73.52}&54.54&53.39&\underline{62.09}&66.59&64.10&\textcolor{blue}{\textbf{84.12}}&57.68&\underline{57.11}&\underline{61.23}&58.14&57.05&\underline{70.24}&\underline{61.81}\\

\rowcolor{uciblue!5}
Qwen3-omni-instruct-30B~\cite{xu2025qwen3}
& -- &15.69&16.88&23.51&22.95&24.06&27.33&40.83&40.76&36.28&18.98&21.81&14.11&24.61&25.88&25.31&25.27 \\

Audio Flamingo 3~\cite{goel2025audio}
& -- &0.00&0.00&0.00&0.00&0.00&0.00&0.00&0.00&0.00&0.66&1.31&1.10&0.17&0.33&0.28&0.26 \\

\rowcolor{uciblue!5}
Qwen2.5-omni~\cite{qwen2025qwen25}
& -- &30.66&33.74&32.07&23.55&23.79&20.79&18.79&19.80&18.30&23.29&24.64&19.13&24.07&25.49&22.57&24.05 \\

\rowcolor{uciblue!15}
\Mname
& -- &\textcolor{blue}{\textbf{72.77}}&\textcolor{blue}{\textbf{75.30}}&\textcolor{blue}{\textbf{78.08}}&\textcolor{blue}{\textbf{75.02}}&\textcolor{blue}{\textbf{73.32}}&\textcolor{blue}{\textbf{71.61}}&\textcolor{blue}{\textbf{79.88}}&\textcolor{blue}{\textbf{78.79}}&78.64&\textcolor{blue}{\textbf{68.91}}&\textcolor{blue}{\textbf{67.19}}&\textcolor{blue}{\textbf{61.67}}&\textcolor{blue}{\textbf{74.15}}&\textcolor{blue}{\textbf{73.65}}&\textcolor{blue}{\textbf{72.50}}&\textcolor{blue}{\textbf{73.43}}\\

\bottomrule\bottomrule
\end{tabular}
}
\caption{V-measure comparison across four corpora.
The best result in each column is highlighted in \textbf{\textcolor{blue}{bold blue}}, and the second-best result is \underline{underlined}.}
\vspace{-1.2em}
\label{tab:V-measure}
\end{table*}

\paragraph{Baselines} We compare our models with both previous paradigms and with different embedding-based clustering approaches. 
We use Whisper~\cite{radford2023robust} as our standard ASR model to transcribe the audio.
For three-step methods, we evaluate general-purpose embedding models (e.g., Qwen3-Embedding~\cite{zhang2025qwen3}, all-MiniLM-L6-v2~\cite{wang2020minilm}) and instruction-tuned embedding models (e.g., InBedder~\cite{peng2024answer}, Instructor~\cite{su2023one}), each combined with classical clustering algorithms K-Means and GMMs.
For instruction-tuned embedding models, we prepend the natural-language clustering perspective as the embedding instruction. For general-purpose embedding models, we encode the ASR transcript directly, as they do not support instruction-conditioned encoding.
For two-step approaches, we use GPT-4o~\cite{hurst2024gpt} as the LLM for clustering.
We also compare \mname with off-the-shelf native LALMs/multimodal large language models (MLLMs): GPT-4o-audio-preview~\cite{openai_gpt4o_audio_preview}, GPT-audio-1.5~\cite{openai_gpt_audio_1_5}, Qwen3-omni-instruct-30B~\cite{xu2025qwen3}, Audio Flamingo 3~\cite{goel2025audio}, Qwen2.5-omni~\cite{qwen2025qwen25}.
For K-Means and GMM initialization, we provide the gold number of clusters to avoid confounding representation quality with cluster-number estimation. This gives these baselines an oracle advantage; in contrast, LLM/LALM-based methods must infer both the number of clusters and assignments from the input. 
More details are provided in App.~\ref{app:baselines}.

\paragraph{Implementation Details}
We adopt two metrics, i.e., V-Measure and Adjusted Rand Index (ARI), to evaluate the model performance, following established practice in clustering
assessment~\cite{cheng2023improving, tipirneni2024context, liu2025llm}. 
The formal definition of the metrics is included in App.~\ref{app:metrics}.
We adopt Audio Flamingo 3~\cite{goel2025audio} as the base model due to its strong understanding and reasoning capabilities acquired during pre-training.
The experiments are conducted on 4 NVIDIA H200 GPUs using the \texttt{TRL} library.
App.~\ref{app:config} provides more details.

\subsection{Main Results}
\label{subsec:main_results}

Tables~\ref{tab:ARI} and~\ref{tab:V-measure} compare \mname against three categories of baselines: ASR-based text embedding followed by conventional clustering, ASR-based LLM clustering, and off-the-shelf native LALMs. 
Overall, \mname achieves the best performance on both evaluation metrics, obtaining an overall ARI of $44.77$ and an overall V-measure of $73.43$. 
Notably, \mname obtains the best ARI on all clustering settings and the best V-measure on $11$ out of $12$ settings. 
The gains are especially pronounced on ECHR, S\&P 500, and Banking77, where \mname consistently achieves the top results across most clustering conditions.
Compared with the strongest competing system, GPT-audio-1.5~\cite{openai_gpt_audio_1_5}, 
\mname improves the overall ARI by $+12.99$ absolute points. 
On V-measure, \mname further surpasses the best baseline by $+11.62$ absolute points.
These consistent improvements across two complementary clustering metrics demonstrate that \mname not only recovers more accurate pairwise grouping structures, but also produces cluster assignments with better homogeneity and completeness. 
Additional results with Qwen2.5-Omni and failure analysis of Audio Flamingo 3 are provided in App.~\ref{app:more_analysis}.

Overall, the results validate the effectiveness of \mname for audio multi-perspective clustering. The large margin over text-embedding baselines confirms the limitation of separating representation learning from clustering, while the improvement over ASR+LLM suggests that directly modeling speech can be beneficial for perspective-conditioned clustering, especially when the criterion depends on paralinguistic cues that may be lost in transcription. More importantly, the consistent gains over strong proprietary LALMs show that task-specific post-training enables \mname to acquire clustering-oriented reasoning and decision-making capabilities beyond those of general-purpose audio-language models.

\begin{figure}[t]
    \centering

    % ========================================================
    % (a) Macro-average ARI
    % ========================================================
    \begin{minipage}[t]{0.48\textwidth}
        \centering
        \resizebox{0.86\linewidth}{!}{%
            \begin{tikzpicture}
                \begin{polaraxis}[
                    width=5.4cm,
                    height=5.4cm,
                    scale only axis,
                    axis equal image,
                    clip=false,
                    ymin=0,
                    ymax=60,
                    ytick={0,15,30,45,60},
                    xtick={0,72,144,216,288},
                    xticklabel=\empty,
                    y tick label style={
                        font=\scriptsize
                    },
                    grid=both,
                    major grid style={
                        draw=gray!50,
                        line width=0.4pt
                    },
                    axis line style={
                        draw=black,
                        line width=0.6pt
                    },
                    legend to name=perspective-model-legend,
                    legend style={
                        legend columns=1,
                        draw=none,
                        font=\scriptsize,
                        column sep=8pt
                    }
                ]

                    % GPT-Audio-1.5
                    \addplot+[
                        thick,
                        mark=*,
                        mark size=2.2pt,
                        % blue!75!black
                        color=green!50!black,
                        mark options={
                            fill=green!50!black,
                            draw=green!50!black
                        }
                    ] coordinates {
                        (0,7.525)
                        (72,2.400)
                        (144,47.925)
                        (216,45.500)
                        (288,42.925)
                        (360,7.525)
                    };
                    \addlegendentry{GPT-Audio-1.5}

                    % Whisper + GPT-4o
                    %
                    % The macro-average Emotion ARI is -1.50.
                    % It is clipped to zero in the radar plot.
                    \addplot+[
                        thick,
                        mark=square*,
                        mark size=2.2pt,
                        % orange!85!black
                        color=orange,
                        mark options={
                            fill=orange,
                            draw=orange
                        }
                    ] coordinates {
                        (0,1.750)
                        (72,0.000)
                        (144,49.850)
                        (216,7.350)
                        (288,41.775)
                        (360,1.750)
                    };
                    \addlegendentry{Whisper + GPT-4o}

                    % Our model
                    \addplot+[
                        very thick,
                        mark=diamond*,
                        mark size=2.5pt,
                        % violet
                        color=uciblue,
                        mark options={
                            fill=uciblue,
                            draw=uciblue
                        }
                    ] coordinates {
                        (0,41.450)
                        (72,39.000)
                        (144,45.520)
                        (216,43.125)
                        (288,48.375)
                        (360,41.450)
                    };
                    \addlegendentry{\Mname}

                    % Perspective labels
                    \node[
                        font=\scriptsize,
                        align=center,
                        anchor=west
                    ] at (axis cs:0,65)
                    {Background\\noise};

                    \node[
                        font=\scriptsize,
                        align=center,
                        anchor=south west
                    ] at (axis cs:72,60)
                    {Emotion};

                    \node[
                        font=\scriptsize,
                        align=center,
                        anchor=south east
                    ] at (axis cs:144,60)
                    {Speaker\\count};

                    \node[
                        font=\scriptsize,
                        align=center,
                        anchor=north east
                    ] at (axis cs:216,60)
                    {Gender};

                    \node[
                        font=\scriptsize,
                        align=center,
                        anchor=north west
                    ] at (axis cs:288,60)
                    {Linguistic\\reasoning};

                \end{polaraxis}
            \end{tikzpicture}%
        }

        \vspace{0.2em}
        % \textbf{(a) Macro-average ARI}
        (a) Perspective-level ARI
    \end{minipage}
    \hfill
    % ========================================================
    % (b) Macro-average V-measure
    % ========================================================
    \begin{minipage}[t]{0.48\textwidth}
        \centering
        \resizebox{0.86\linewidth}{!}{%
            \begin{tikzpicture}
                \begin{polaraxis}[
                    width=5.4cm,
                    height=5.4cm,
                    scale only axis,
                    axis equal image,
                    clip=false,
                    ymin=0,
                    ymax=80,
                    ytick={0,20,40,60,80},
                    xtick={0,72,144,216,288},
                    xticklabel=\empty,
                    y tick label style={
                        font=\scriptsize
                    },
                    grid=both,
                    major grid style={
                        draw=gray!50,
                        line width=0.4pt
                    },
                    axis line style={
                        draw=black,
                        line width=0.6pt
                    }
                ]

                    % GPT-Audio-1.5
                    \addplot+[
                        thick,
                        mark=*,
                        mark size=2.2pt,
                        % blue!75!black
                        color=green!50!black,
                        mark options={
                            fill=green!50!black,
                            draw=green!50!black
                        }
                    ] coordinates {
                        (0,24.850)
                        (72,36.200)
                        (144,65.175)
                        (216,63.900)
                        (288,74.200)
                        (360,24.850)
                    };

                    % Whisper + GPT-4o
                    \addplot+[
                        thick,
                        mark=square*,
                        mark size=2.2pt,
                        % orange!85!black
                        color=orange,
                        mark options={
                            fill=orange,
                            draw=orange
                        }
                    ] coordinates {
                        (0,32.775)
                        (72,37.625)
                        (144,59.325)
                        (216,27.325)
                        (288,75.150)
                        (360,32.775)
                    };

                    % Our model
                    \addplot+[
                        very thick,
                        mark=diamond*,
                        mark size=2.5pt,
                        % violet
                        color=uciblue,
                        mark options={
                            fill=uciblue,
                            draw=uciblue
                        }
                    ] coordinates {
                        (0,60.350)
                        (72,69.000)
                        (144,58.215)
                        (216,54.250)
                        (288,77.100)
                        (360,60.350)
                    };

                    % Perspective labels
                    \node[
                        font=\scriptsize,
                        align=center,
                        anchor=west
                    ] at (axis cs:0,85)
                    {Background\\noise};

                    \node[
                        font=\scriptsize,
                        align=center,
                        anchor=south west
                    ] at (axis cs:72,80)
                    {Emotion};

                    \node[
                        font=\scriptsize,
                        align=center,
                        anchor=south east
                    ] at (axis cs:144,80)
                    {Speaker\\count};

                    \node[
                        font=\scriptsize,
                        align=center,
                        anchor=north east
                    ] at (axis cs:216,80)
                    {Gender};

                    \node[
                        font=\scriptsize,
                        align=center,
                        anchor=north west
                    ] at (axis cs:288,80)
                    {Linguistic\\reasoning};

                \end{polaraxis}
            \end{tikzpicture}%
        }

        \vspace{0.2em}
        % \textbf{(b) Macro-average V-measure}
        (b) Perspective-level V-measure
    \end{minipage}

    \vspace{0.3em}

    \pgfplotslegendfromname{perspective-model-legend}

    \vspace{-0.3em}

    \caption{
        Perspective-level performance averaged across the corpora. 
        Each axis represents one clustering perspective, and each curve represents one model.
        Because polar coordinates cannot faithfully represent a negative radius, the ARI of \(-1.50\) obtained
        by Whisper + GPT-4o for emotion clustering is displayed at zero.
        All other points show their exact values.
    }
    \label{fig:perspective-level-radar}
\end{figure}

\subsection{Discussion}

\begin{table}[t]
\centering
\small
\setlength{\tabcolsep}{4.2pt}
\renewcommand{\arraystretch}{1.08}
\resizebox{\columnwidth}{!}{
\begin{tabular}{llcccc}
\toprule
\toprule
Metric & Corpus / Split &
\shortstack{Answer-only\\SFT} &
\shortstack{RD\\SFT} &
\shortstack{Answer-only\\+ DPO} &
\shortstack{RD\\+ DPO} \\
\midrule

\multirow{6}{*}{ARI}
& \cellcolor{uciblue!5}ECHR       
& \cellcolor{uciblue!5}33.17 
& \cellcolor{uciblue!5}33.85 
& \cellcolor{uciblue!5}35.38 
& \cellcolor{uciblue!5}\textcolor{blue}{\textbf{42.02}} \\

& S\&P 500   
& 33.30 
& 38.23 
& 37.58 
& \textcolor{blue}{\textbf{42.73}} \\

& \cellcolor{uciblue!5}Banking77  
& \cellcolor{uciblue!5}39.50 
& \cellcolor{uciblue!5}39.48 
& \cellcolor{uciblue!5}41.26 
& \cellcolor{uciblue!5}\textcolor{blue}{\textbf{49.85}} \\

& MultiWOZ   
& 33.34 
& 32.34 
& 34.07 
& \textcolor{blue}{\textbf{44.50}} \\

& \cellcolor{uciblue!5}Overall    
& \cellcolor{uciblue!5}34.83 
& \cellcolor{uciblue!5}35.97 
& \cellcolor{uciblue!5}37.07 
& \cellcolor{uciblue!5}\textcolor{blue}{\textbf{44.77}} \\

& \cellcolor{uciblue!15}$\Delta$   
& \cellcolor{uciblue!15}--    
& \cellcolor{uciblue!15}$+1.14$ 
& \cellcolor{uciblue!15}$+2.24$ 
& \cellcolor{uciblue!15}\textcolor{blue}{$\mathbf{+9.94}$} \\
\midrule

\multirow{6}{*}{V-measure}
& \cellcolor{uciblue!5}ECHR       
& \cellcolor{uciblue!5}69.30 
& \cellcolor{uciblue!5}69.20 
& \cellcolor{uciblue!5}70.96 
& \cellcolor{uciblue!5}\textcolor{blue}{\textbf{75.38}} \\

& S\&P 500   
& 66.27 
& 69.88 
& 69.90 
& \textcolor{blue}{\textbf{73.32}} \\

& \cellcolor{uciblue!5}Banking77  
& \cellcolor{uciblue!5}71.21 
& \cellcolor{uciblue!5}72.51 
& \cellcolor{uciblue!5}74.13 
& \cellcolor{uciblue!5}\textcolor{blue}{\textbf{79.10}} \\

& MultiWOZ   
& 55.94 
& 52.26 
& 54.56 
& \textcolor{blue}{\textbf{65.92}} \\

& \cellcolor{uciblue!5}Overall    
& \cellcolor{uciblue!5}65.68 
& \cellcolor{uciblue!5}65.96 
& \cellcolor{uciblue!5}67.39 
& \cellcolor{uciblue!5}\textcolor{blue}{\textbf{73.43}} \\

& \cellcolor{uciblue!15}$\Delta$   
& \cellcolor{uciblue!15}--    
& \cellcolor{uciblue!15}$+0.28$ 
& \cellcolor{uciblue!15}$+1.71$ 
& \cellcolor{uciblue!15}\textcolor{blue}{$\mathbf{+7.75}$} \\

\bottomrule
\bottomrule
\end{tabular}
}
\vspace{-.5em}
\caption{
Ablation study of answer-only supervised fine-tuning (SFT), RD, and DPO.
RD SFT denotes supervised fine-tuning with RD, while Answer-only DPO and RD+DPO denote DPO initialized from the corresponding checkpoints.
$\Delta$ denotes the absolute improvement over the answer-only SFT baseline.
}
\vspace{-2em}
\label{tab:ablation_cot_dpo}
\end{table}

\paragraph{Perspective-level Performance.}
The corpus-macro-averaged results are shown in Fig.~\ref{fig:perspective-level-radar}. \mname achieves the strongest average performance on background-noise, emotion, and linguistic-reasoning perspectives under both ARI and V-measure. 
The improvements are particularly obvious for background noise and emotion: \mname obtains (41.45/60.35) and (39.00/69.00), respectively, compared with the strongest corresponding baseline results of (7.53/32.78) and (2.40/37.63). 
On linguistic reasoning, \mname also achieves the best macro-average result of (48.38/77.10). 
These results indicate that the gains do not arise solely from lexical-semantic reasoning, but extend to perspectives that require direct modeling of acoustic and paralinguistic information.

The baselines exhibit more specialized behavior. Whisper+GPT-4o remains competitive on linguistic reasoning, where transcripts preserve much of the relevant clustering signal, but performs substantially worse on background noise, emotion, and gender. 
The performance of GPT-Audio-1.5 is considerably stronger but nevertheless uneven across perspectives, particularly for background noise and emotion. 
In contrast, \mname exhibits a substantially more balanced profile: its average ARI ranges from 36.13 to 48.38 across the five perspectives, while its V-measure ranges from 54.25 to 77.10. 
This consistency suggests broader perspective-conditioned clustering ability rather than specialization toward either text-dominant or narrowly defined acoustic criteria. 

\paragraph{Ablation Study}Tab.~\ref{tab:ablation_cot_dpo} studies the contribution of RD and DPO. The first row, which uses answer-only supervised fine-tuning (SFT) without RD or DPO, serves as the baseline. We observe that applying RD alone yields a modest improvement in overall ARI, increasing from 34.83 to 35.97, with a $+1.14$ absolute gain. However, its effect on V-measure is relatively small, improving the overall score by only $+0.28$. This suggests that RD can provide useful supervision for improving clustering assignments, but by itself, it does not consistently lead to better cluster quality across all corpora.

DPO alone provides a stronger gain than RD alone. Compared with the baseline, DPO improves the overall ARI from 34.83 to 37.07, corresponding to a $+2.24$ absolute improvement, and improves the overall V-measure from 65.68 to 67.39, with a $+1.71$ gain. The improvement is especially clear on ECHR, S\&P 500, and Banking77, indicating that preference optimization helps the model better align its outputs with desirable clustering structures. Nevertheless, DPO alone still shows limited improvement in some cases, such as MultiWOZ in V-measure, suggesting that preference learning without additional reasoning supervision may not fully resolve more challenging ambiguities.

The best performance is obtained when RD and DPO are combined. This setting achieves 44.77 overall ARI and 73.43 overall V-measure, outperforming the baseline by $+9.94$ and $+7.75$ absolute points, respectively. Importantly, the combined setting improves consistently across all four corpora for both metrics. Compared with DPO alone, adding RD further improves overall ARI by $+7.70$ and V-measure by $+6.04$, showing that RD and DPO are complementary rather than redundant. These results suggest that RD provides a stronger intermediate reasoning signal, while DPO further calibrates the model toward preferred clustering decisions. Together, they substantially improve both pairwise clustering agreement and cluster-level homogeneity/completeness, demonstrating the effectiveness of combining reasoning-based supervision with preference optimization.

\begin{figure}[tb]
\centering

\begin{subfigure}{0.48\columnwidth}
    \centering
    \includegraphics[width=\linewidth]
    {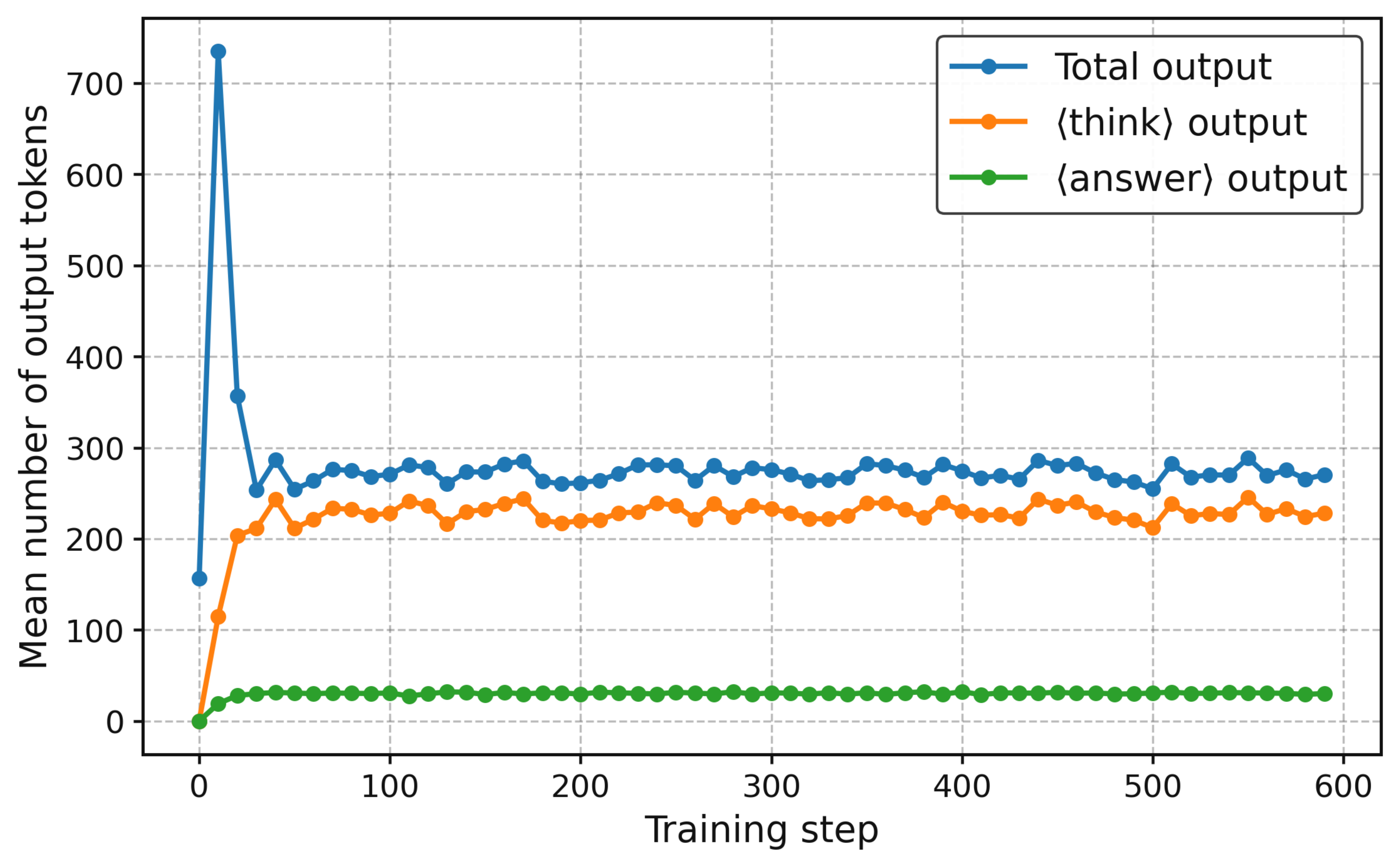}
    \caption{Response length decomposition.}
    \label{fig:output_length}
\end{subfigure}
\hfill
\begin{subfigure}{0.48\columnwidth}
    \centering
    \includegraphics[width=\linewidth]{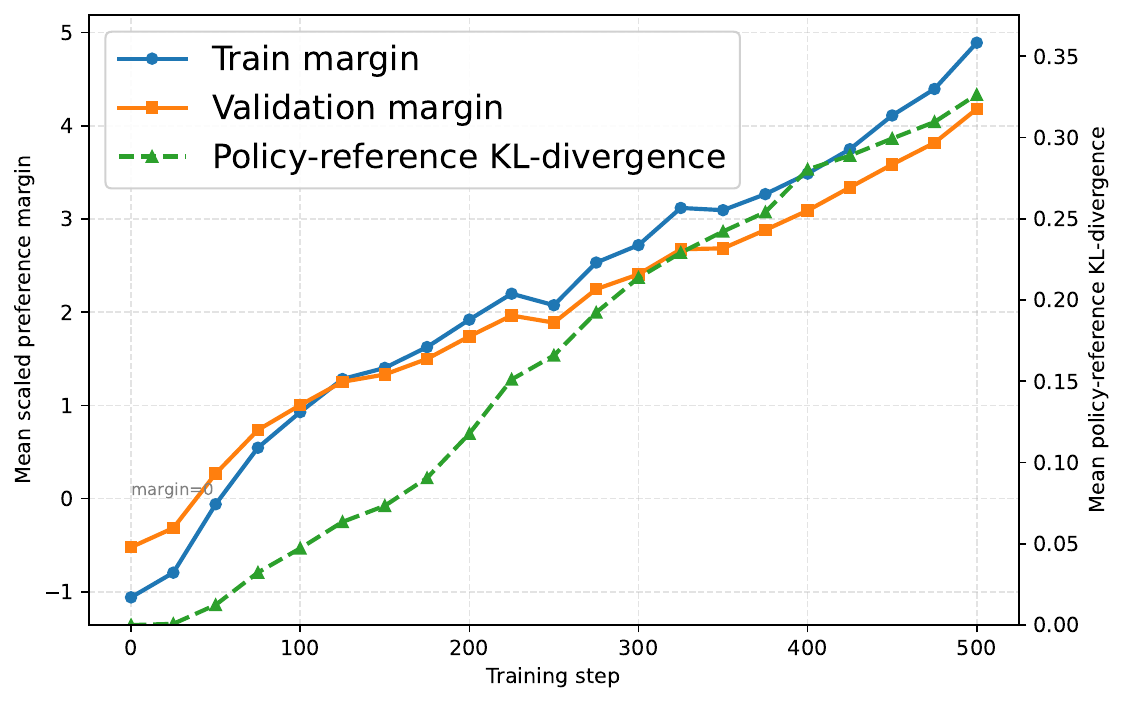}
    \caption{Preference margin and KL divergence.}
    \label{fig:margin_kl_combined}
\end{subfigure}
\vspace{-1em}
\caption{
(a) Evolution of mean output length, decomposed into total output tokens, \thinktag{} tokens, and \answertag{} tokens. (b) Evolution of the scaled preference margin and policy-reference KL divergence.
The margin is computed as
$m_{\mathrm{scaled}}
=
L_{\mathrm{ref}}
\left(
\overline{\log \pi_{\theta}(y^{+}\mid x)}
-
\overline{\log \pi_{\theta}(y^{-}\mid x)}
\right)$.
}
\label{fig:resp_kl}
\vspace{-2em}
\end{figure}

\paragraph{Response-format Stabilization} Fig.~\ref{fig:resp_kl}(a) shows the mean number of generated tokens across training steps, with separate measurements for the complete model output and for tokens enclosed within the prescribed \thinktag{} and \answertag{} fields. At the beginning of training, the total output length exhibits a sharp transient spike, whereas the counted \thinktag{} and \answertag{} tokens remain substantially lower. 
This discrepancy indicates that the model initially fails to consistently follow the required response format, producing a large number of extraneous tokens outside the designated fields. 
After this short adaptation phase, the three curves stabilize: the \thinktag{} segment accounts for the majority of the response, and the \answertag{} segment remains compact and stable.
These trends suggest that training rapidly improves structural adherence and suppresses uncontrolled generation, leading to a consistent output format throughout the remaining optimization process.

\paragraph{Preference Alignment and Controlled Policy Shift} Fig.~\ref{fig:resp_kl}(b) shows that DPO training steadily increases the model's preference margin while also inducing a gradual divergence from the reference model.
At the beginning, both training and validation margins are negative, indicating that the policy does not yet assign a higher likelihood to the chosen responses than to the rejected ones.
As training proceeds, both margins rapidly cross the zero-margin boundary and continue to increase, suggesting that DPO effectively shifts the policy toward preference-aligned outputs.
The validation margin closely tracks the training margin across training, with only a moderate gap emerging in later stages, which suggests that the learned preference signal transfers beyond the training probe rather than merely memorizing the optimization examples.
Meanwhile, the policy-reference KL-divergence increases smoothly from near zero to a moderate value, indicating that the policy gradually departs from the reference model as preference optimization progresses.
The KL curve grows in a controlled manner rather than exhibiting abrupt spikes, suggesting that the improvement is achieved through a stable distributional shift.

\section{Related Work}

\paragraph{Traditional and Embedding-based Clustering}

Clustering is a fundamental problem in machine learning, with classical methods such as K-Means~\cite{macqueen1967some, lloyd1982least} and GMMs~\cite{dempster1977maximum} widely used to partition data. 
Recent representation learning methods improve clustering by using pretrained encoders. In speech processing, 
wav2vec 2.0~\cite{baevski2020wav2vec} and HuBERT~\cite{hsu2021hubert} learn rich acoustic representations. Instruction-aware embedding models, including Instructor~\cite{su2023one} and InBedder~\cite{peng2024answer}, further condition representations on natural-language instructions. 

\paragraph{LLMs and Reasoning-based Clustering}

LLMs have recently been explored as components in clustering systems~\cite{zhang2023clusterllm, viswanathan2024large, deraedt2023idas, feng2024llmedgerefine}. More recently, reasoning-based clustering methods such as Cluster-R1~\cite{qing2026cluster} formulate clustering as a generative reasoning task. These methods show the potential of using LLM reasoning for flexible structure discovery.

\paragraph{Audio-Language Models and Multimodal Reasoning}

Recent LALMs, such as SpeechT5~\cite{ao2022speecht5}, AudioLM~\cite{borsos2023audiolm}, and Qwen2-Audio~\cite{qwen2024audio}, enable end-to-end modeling of speech and text within unified frameworks.  
Despite this progress, prior work has largely focused on speech recognition, generation, or audio understanding, leaving structure discovery over speech collections underexplored. Our work addresses this gap by studying audio multi-perspective clustering with end-to-end audio-language models.

\section{Conclusion}
We introduced audio multi-perspective clustering, a new task that requires models to organize a collection of speech recordings according to a natural-language perspective while inferring both the number of clusters and the cluster assignments. To support this task, we constructed \Dname, a benchmark spanning diverse application domains and combining linguistic and paralinguistic clustering perspectives. We further proposed \Mname, an end-to-end LALM trained with reasoning distillation and preference optimization to perform perspective-conditioned structure discovery directly from speech. Experiments show that \Mname{} consistently outperforms the baselines across ARI and V-measure. 
These results suggest that native LALMs can serve as flexible clustering agents for speech collections, opening new directions for perspective-conditioned organization, retrieval, and analysis of audio data.

\section*{Author Contributions}
Wenjun Huang led the project from conception to completion, including problem formulation and idea framing, benchmark design and construction, methodology development, experimental design and implementation. Wenjun Huang also coordinated the overall research workflow, led the writing, revision, and finalization of the manuscript.
Qiaosong Chu designed and implemented the core reusable benchmark construction and evaluation pipeline, including data processing, TTS generation, benchmark sampling, and evaluation scripts; instantiated the pipeline on the ECHR and SP500 datasets; implemented the SFT/RL training and evaluation framework; and conducted the main model training, model evaluation, and RL experiments.

\bibliography{custom}

\begin{thebibliography}{39}
\providecommand{\natexlab}[1]{#1}

\bibitem[{Ao et~al.(2022)Ao, Wang, Zhou, Wang, Ren, Wu, Liu, Ko, Li, Zhang et~al.}]{ao2022speecht5}
Junyi Ao, Rui Wang, Long Zhou, Chengyi Wang, Shuo Ren, Yu~Wu, Shujie Liu, Tom Ko, Qing Li, Yu~Zhang, and 1 others. 2022.
\newblock Speecht5: Unified-modal encoder-decoder pre-training for spoken language processing.
\newblock In \emph{Annual Meeting of the Association for Computational Linguistics}.
\newblock Available: \url{https://aclanthology.org/2022.acl-long.393.pdf}.

\bibitem[{Baevski et~al.(2020)Baevski, Zhou, Mohamed, and Auli}]{baevski2020wav2vec}
Alexei Baevski, Yuhao Zhou, Abdelrahman Mohamed, and Michael Auli. 2020.
\newblock wav2vec 2.0: A framework for self-supervised learning of speech representations.
\newblock In \emph{Advances in Neural Information Processing Systems}.
\newblock Available: \url{https://arxiv.org/pdf/2006.11477}.

\bibitem[{Borsos et~al.(2023)Borsos, Marinier, Vincent, Kharitonov, Pietquin, Sharifi, Roblek, Teboul, Grangier, Tagliasacchi et~al.}]{borsos2023audiolm}
Zal{\'a}n Borsos, Rapha{\"e}l Marinier, Damien Vincent, Eugene Kharitonov, Olivier Pietquin, Matt Sharifi, Dominik Roblek, Olivier Teboul, David Grangier, Marco Tagliasacchi, and 1 others. 2023.
\newblock Audiolm: a language modeling approach to audio generation.
\newblock \emph{Transactions on Audio, Speech, and Language Processing}.
\newblock Available: \url{https://arxiv.org/pdf/2209.03143}.

\bibitem[{Casanueva et~al.(2020)Casanueva, Tem{\v{c}}inas, Gerz, Henderson, and Vuli{\'c}}]{casanueva2020efficient}
I{\~n}igo Casanueva, Tadas Tem{\v{c}}inas, Daniela Gerz, Matthew Henderson, and Ivan Vuli{\'c}. 2020.
\newblock Efficient intent detection with dual sentence encoders.
\newblock In \emph{2nd Workshop on Natural Language Processing for Conversational AI}.
\newblock Available: \url{https://aclanthology.org/2020.nlp4convai-1.5.pdf}.

\bibitem[{Cheng et~al.(2023)Cheng, Zheng, Zhang, Wang, Chen, Chen, and Zhang}]{cheng2023improving}
Luyao Cheng, Siqi Zheng, Qinglin Zhang, Hui Wang, Yafeng Chen, Qian Chen, and Shiliang Zhang. 2023.
\newblock Improving speaker diarization using semantic information: joint pairwise constraints propagation.
\newblock \emph{arXiv preprint arXiv:2309.10456}.
\newblock Available: \url{https://arxiv.org/pdf/2309.10456}.

\bibitem[{Chu et~al.(2024)Chu, Xu, Yang, Wei, Wei, Guo, Leng, Lv, He, Lin et~al.}]{qwen2024audio}
Yunfei Chu, Jin Xu, Qian Yang, Haojie Wei, Xipin Wei, Zhifang Guo, Yichong Leng, Yuanjun Lv, Jinzheng He, Junyang Lin, and 1 others. 2024.
\newblock Qwen2-audio technical report.
\newblock \emph{arXiv preprint arXiv:2407.10759}.
\newblock Available: \url{http://arxiv.org/pdf/2407.10759}.

\bibitem[{Clifton et~al.(2020)Clifton, Reddy, Yu, Pappu, Rezapour, Bonab, Eskevich, Jones, Karlgren, Carterette et~al.}]{clifton2020podcasts}
Ann Clifton, Sravana Reddy, Yongze Yu, Aasish Pappu, Rezvaneh Rezapour, Hamed Bonab, Maria Eskevich, Gareth Jones, Jussi Karlgren, Ben Carterette, and 1 others. 2020.
\newblock 100,000 podcasts: A spoken english document corpus.
\newblock In \emph{International Conference on Computational Linguistics}.
\newblock Available: \url{https://aclanthology.org/2020.coling-main.519.pdf}.

\bibitem[{De~Raedt et~al.(2023)De~Raedt, Godin, Demeester, and Develder}]{deraedt2023idas}
Maarten De~Raedt, Fr{\'e}deric Godin, Thomas Demeester, and Chris Develder. 2023.
\newblock Idas: Intent discovery with abstractive summarization.
\newblock In \emph{5th Workshop on NLP for Conversational AI}.
\newblock Available: \url{https://aclanthology.org/2023.nlp4convai-1.7.pdf}.

\bibitem[{Dempster et~al.(1977)Dempster, Laird, and Rubin}]{dempster1977maximum}
Arthur~P Dempster, Nan~M Laird, and Donald~B Rubin. 1977.
\newblock Maximum likelihood from incomplete data via the em algorithm.
\newblock \emph{Journal of the Royal Statistical Society: Series B (Methodological)}.
\newblock Available: \url{https://www.ece.iastate.edu/~namrata/EE527_Spring08/Dempster77.pdf}.

\bibitem[{Diao et~al.(2025)Diao, Zhang, Kong, Wu, Ma, Ouyang, Qing, Vosoughi, and Gui}]{diao2025soundmind}
Xingjian Diao, Chunhui Zhang, Keyi Kong, Weiyi Wu, Chiyu Ma, Zhongyu Ouyang, Peijun Qing, Soroush Vosoughi, and Jiang Gui. 2025.
\newblock Soundmind: Rl-incentivized logic reasoning for audio-language models.
\newblock In \emph{Conference on Empirical Methods in Natural Language Processing}.
\newblock Available: \url{https://aclanthology.org/2025.emnlp-main.27.pdf}.

\bibitem[{Feng et~al.(2024)Feng, Lin, Wang, Cheng, and Wong}]{feng2024llmedgerefine}
Zijin Feng, Luyang Lin, Lingzhi Wang, Hong Cheng, and Kam-Fai Wong. 2024.
\newblock Llmedgerefine: Enhancing text clustering with llm-based boundary point refinement.
\newblock In \emph{Conference on Empirical Methods in Natural Language Processing}.
\newblock Available: \url{https://aclanthology.org/2024.emnlp-main.1025.pdf}.

\bibitem[{Goel et~al.(2025)Goel, Ghosh, Kim, Kumar, Kong, Lee, Yang, Duraiswami, Manocha, Valle et~al.}]{goel2025audio}
Arushi Goel, Sreyan Ghosh, Jaehyeon Kim, Sonal Kumar, Zhifeng Kong, Sang-gil Lee, Chao-Han~Huck Yang, Ramani Duraiswami, Dinesh Manocha, Rafael Valle, and 1 others. 2025.
\newblock Audio flamingo 3: Advancing audio intelligence with fully open large audio language models.
\newblock \emph{arXiv preprint arXiv:2507.08128}.
\newblock Available: \url{https://arxiv.org/pdf/2507.08128}.

\bibitem[{Hsu et~al.(2021)Hsu, Bolte, Tsai, Lakhotia, Salakhutdinov, and Mohamed}]{hsu2021hubert}
Wei-Ning Hsu, Benjamin Bolte, Yao-Hung~Hubert Tsai, Kushal Lakhotia, Ruslan Salakhutdinov, and Abdelrahman Mohamed. 2021.
\newblock Hubert: Self-supervised speech representation learning by masked prediction of hidden units.
\newblock \emph{Transactions on Audio, Speech, and Language Processing}.
\newblock Available: \url{https://arxiv.org/pdf/2106.07447}.

\bibitem[{Hu et~al.(2023)Hu, Ganter, Deilamsalehy, Dernoncourt, Foroosh, and Liu}]{hu2023meetingbank}
Yebowen Hu, Timothy Ganter, Hanieh Deilamsalehy, Franck Dernoncourt, Hassan Foroosh, and Fei Liu. 2023.
\newblock Meetingbank: A benchmark dataset for meeting summarization.
\newblock In \emph{Annual Meeting of the Association for Computational Linguistics}.
\newblock Available: \url{https://aclanthology.org/2023.acl-long.906.pdf}.

\bibitem[{Hubert and Arabie(1985)}]{hubert1985comparing}
Lawrence Hubert and Phipps Arabie. 1985.
\newblock Comparing partitions.
\newblock \emph{Journal of Classification}.
\newblock Available: \url{https://link.springer.com/article/10.1007/BF01908075}.

\bibitem[{Hurst et~al.(2024)Hurst, Lerer, Goucher, Perelman, Ramesh, Clark, Ostrow, Welihinda, Hayes, Radford et~al.}]{hurst2024gpt}
Aaron Hurst, Adam Lerer, Adam~P Goucher, Adam Perelman, Aditya Ramesh, Aidan Clark, AJ~Ostrow, Akila Welihinda, Alan Hayes, Alec Radford, and 1 others. 2024.
\newblock Gpt-4o system card.
\newblock \emph{arXiv preprint arXiv:2410.21276}.
\newblock Available: \url{https://arxiv.org/pdf/2410.21276}.

\bibitem[{jlohding(2023)}]{jlohding_sp500_edgar_10k}
jlohding. 2023.
\newblock Sp500-edgar-10k: Annual reports of s\&p 500 companies from sec filings.
\newblock Available: \url{https://huggingface.co/datasets/jlohding/sp500-edgar-10k}.

\bibitem[{Larson and Jones(2012)}]{larson2012spoken}
Martha Larson and Gareth~JF Jones. 2012.
\newblock Spoken content retrieval: A survey of techniques and technologies.
\newblock \emph{Foundations and Trends{\textregistered} in Information Retrieval}.
\newblock Available: \url{https://www.emerald.com/ftinr/article-pdf/5/4-5/235/11085594/1500000020en.pdf}.

\bibitem[{Liu et~al.(2025)Liu, Shang, Ke, Wang, Luo, Liu, Li, and Li}]{liu2025llm}
Jianghan Liu, Ziyu Shang, Wenjun Ke, Peng Wang, Zhizhao Luo, Jiajun Liu, Guozheng Li, and Yining Li. 2025.
\newblock Llm-guided semantic-aware clustering for topic modeling.
\newblock In \emph{Annual Meeting of the Association for Computational Linguistics}.
\newblock Available: \url{https://aclanthology.org/2025.acl-long.902.pdf}.

\bibitem[{Lloyd(1982)}]{lloyd1982least}
Stuart Lloyd. 1982.
\newblock Least squares quantization in pcm.
\newblock \emph{IEEE Transactions on Information Theory}.
\newblock Available: \url{https://ieeexplore.ieee.org/document/1056489}.

\bibitem[{MacQueen(1967)}]{macqueen1967some}
J.~MacQueen. 1967.
\newblock Some methods for classification and analysis of multivariate observations.
\newblock In \emph{Fifth Berkeley Symposium on Mathematical Statistics and Probability}.
\newblock Available: \url{ https://cir.nii.ac.jp/crid/1570572699967325952}.

\bibitem[{{OpenAI}(2025)}]{openai_gpt4o_audio_preview}
{OpenAI}. 2025.
\newblock {GPT-4o Audio Model}.
\newblock Available: \url{https://developers.openai.com/api/docs/models/gpt-4o-audio-preview}.

\bibitem[{{OpenAI}(2026)}]{openai_gpt_audio_1_5}
{OpenAI}. 2026.
\newblock {GPT-Audio 1.5 Model}.
\newblock Available: \url{https://developers.openai.com/api/docs/models/gpt-audio-1.5}.

\bibitem[{Park et~al.(2022)Park, Kanda, Dimitriadis, Han, Watanabe, and Narayanan}]{park2022review}
Tae~Jin Park, Naoyuki Kanda, Dimitrios Dimitriadis, Kyu~J Han, Shinji Watanabe, and Shrikanth Narayanan. 2022.
\newblock A review of speaker diarization: Recent advances with deep learning.
\newblock \emph{Computer Speech \& Language}.
\newblock Available: \url{https://www.sciencedirect.com/science/article/abs/pii/S0885230821001121}.

\bibitem[{Peng et~al.(2024)Peng, Zhang, Wang, Srinivasa, Liu, Wang, and Shang}]{peng2024answer}
Letian Peng, Yuwei Zhang, Zilong Wang, Jayanth Srinivasa, Gaowen Liu, Zihan Wang, and Jingbo Shang. 2024.
\newblock Answer is all you need: Instruction-following text embedding via answering the question.
\newblock In \emph{Annual Meeting of the Association for Computational Linguistics}.
\newblock Available: \url{ https://aclanthology.org/2024.acl-long.27.pdf}.

\bibitem[{Poudyal et~al.(2020)Poudyal, {\v{S}}avelka, Ieven, Moens, Goncalves, and Quaresma}]{poudyal2020echr}
Prakash Poudyal, Jarom{\'\i}r {\v{S}}avelka, Aagje Ieven, Marie~Francine Moens, Teresa Goncalves, and Paulo Quaresma. 2020.
\newblock Echr: Legal corpus for argument mining.
\newblock In \emph{7th Workshop on Argument Mining}.
\newblock Available: \url{https://aclanthology.org/2020.argmining-1.8.pdf}.

\bibitem[{Qing et~al.(2026)Qing, Mathur, Lipka, Manjunatha, Rossi, Dernoncourt, Hassanpour, and Vosoughi}]{qing2026cluster}
Peijun Qing, Puneet Mathur, Nedim Lipka, Varun Manjunatha, Ryan Rossi, Franck Dernoncourt, Saeed Hassanpour, and Soroush Vosoughi. 2026.
\newblock Cluster-r1: Large reasoning models are instruction-following clustering agents.
\newblock \emph{arXiv preprint arXiv:2603.23518}.
\newblock Available: \url{https://arxiv.org/pdf/2603.23518}.

\bibitem[{Qwen(2025)}]{qwen2025qwen25}
Team Qwen. 2025.
\newblock Qwen2.5-omni technical report.
\newblock \emph{arXiv preprint arXiv:2503.20215}.
\newblock Available: \url{https://arxiv.org/pdf/2503.20215}.

\bibitem[{Radford et~al.(2023)Radford, Kim, Xu, Brockman, McLeavey, and Sutskever}]{radford2023robust}
Alec Radford, Jong~Wook Kim, Tao Xu, Greg Brockman, Christine McLeavey, and Ilya Sutskever. 2023.
\newblock Robust speech recognition via large-scale weak supervision.
\newblock In \emph{International Conference on Machine Learning}, pages 28492--28518.
\newblock Available: \url{https://arxiv.org/pdf/2212.04356}.

\bibitem[{Reimers and Gurevych(2019)}]{reimers2019sentence}
Nils Reimers and Iryna Gurevych. 2019.
\newblock Sentence-bert: Sentence embeddings using siamese bert-networks.
\newblock In \emph{Conference on Empirical Methods in Natural Language Processing and the 9th International Joint Conference on Natural Language Processing}.
\newblock Available: \url{https://aclanthology.org/D19-1410.pdf}.

\bibitem[{Rosenberg and Hirschberg(2007)}]{rosenberg2007v}
Andrew Rosenberg and Julia Hirschberg. 2007.
\newblock V-measure: A conditional entropy-based external cluster evaluation measure.
\newblock In \emph{Joint Conference on Empirical Methods in Natural Language Processing and Computational Natural Language Learning}.
\newblock Available: \url{https://aclanthology.org/D07-1043.pdf}.

\bibitem[{Su et~al.(2023)Su, Shi, Kasai, Wang, Hu, Ostendorf, Yih, Smith, Zettlemoyer, and Yu}]{su2023one}
Hongjin Su, Weijia Shi, Jungo Kasai, Yizhong Wang, Yushi Hu, Mari Ostendorf, Wen-tau Yih, Noah~A Smith, Luke Zettlemoyer, and Tao Yu. 2023.
\newblock One embedder, any task: Instruction-finetuned text embeddings.
\newblock In \emph{Findings of the Association for Computational Linguistics: ACL 2023}.
\newblock Available: \url{ https://aclanthology.org/2023.findings-acl.71.pdf}.

\bibitem[{Tipirneni et~al.(2024)Tipirneni, Adkathimar, Choudhary, Hiranandani, Amjad, Ioannidis, Yuan, and Reddy}]{tipirneni2024context}
Sindhu Tipirneni, Ravinarayana Adkathimar, Nurendra Choudhary, Gaurush Hiranandani, Rana~Ali Amjad, Vassilis~N Ioannidis, Changhe Yuan, and Chandan~K Reddy. 2024.
\newblock Context-aware clustering using large language models.
\newblock \emph{arXiv preprint arXiv:2405.00988}.
\newblock Available: \url{https://arxiv.org/pdf/2405.00988}.

\bibitem[{Viswanathan et~al.(2024)Viswanathan, Gashteovski, Lawrence, Wu, and Neubig}]{viswanathan2024large}
Vijay Viswanathan, Kiril Gashteovski, Carolin Lawrence, Tongshuang Wu, and Graham Neubig. 2024.
\newblock Large language models enable few-shot clustering.
\newblock \emph{Transactions of the Association for Computational Linguistics}.
\newblock Available: \url{https://aclanthology.org/2024.tacl-1.18.pdf}.

\bibitem[{Wang et~al.(2020)Wang, Wei, Dong, Bao, Yang, and Zhou}]{wang2020minilm}
Wenhui Wang, Furu Wei, Li~Dong, Hangbo Bao, Nan Yang, and Ming Zhou. 2020.
\newblock Minilm: Deep self-attention distillation for task-agnostic compression of pre-trained transformers.
\newblock In \emph{Advances in Neural Information Processing Systems}.
\newblock Available: \url{https://proceedings.neurips.cc/paper/2020/file/3f5ee243547dee91fbd053c1c4a845aa-Paper.pdf}.

\bibitem[{Xu et~al.(2025)Xu, Guo, Hu, Chu, Wang, He, Wang, Shi, He, Zhu et~al.}]{xu2025qwen3}
Jin Xu, Zhifang Guo, Hangrui Hu, Yunfei Chu, Xiong Wang, Jinzheng He, Yuxuan Wang, Xian Shi, Ting He, Xinfa Zhu, and 1 others. 2025.
\newblock Qwen3-omni technical report.
\newblock \emph{arXiv preprint arXiv:2509.17765}.
\newblock Available: \url{https://arxiv.org/pdf/2509.17765}.

\bibitem[{Zang et~al.(2020)Zang, Rastogi, Sunkara, Gupta, Zhang, and Chen}]{zang2020multiwoz}
Xiaoxue Zang, Abhinav Rastogi, Srinivas Sunkara, Raghav Gupta, Jianguo Zhang, and Jindong Chen. 2020.
\newblock Multiwoz 2.2: A dialogue dataset with additional annotation corrections and state tracking baselines.
\newblock In \emph{2nd Workshop on Natural Language Processing for Conversational AI}.
\newblock Available: \url{https://aclanthology.org/2020.nlp4convai-1.13.pdf}.

\bibitem[{Zhang et~al.(2025)Zhang, Li, Long, Zhang, Lin, Yang, Xie, Yang, Liu, Lin et~al.}]{zhang2025qwen3}
Yanzhao Zhang, Mingxin Li, Dingkun Long, Xin Zhang, Huan Lin, Baosong Yang, Pengjun Xie, An~Yang, Dayiheng Liu, Junyang Lin, and 1 others. 2025.
\newblock Qwen3 embedding: Advancing text embedding and reranking through foundation models.
\newblock \emph{arXiv preprint arXiv:2506.05176}.
\newblock Available: \url{https://arxiv.org/pdf/2506.05176}.

\bibitem[{Zhang et~al.(2023)Zhang, Wang, and Shang}]{zhang2023clusterllm}
Yuwei Zhang, Zihan Wang, and Jingbo Shang. 2023.
\newblock Clusterllm: Large language models as a guide for text clustering.
\newblock In \emph{Conference on Empirical Methods in Natural Language Processing}.
\newblock Available: \url{https://aclanthology.org/2023.emnlp-main.858.pdf}.

\end{thebibliography}

\appendix

\startcontents[appendices]
\section*{Contents of Appendix}
\printcontents[appendices]{}{1}{\normalsize}

\section{Corpus-Specific Benchmark Construction}
\label{app:corpus_pipeline}

This appendix provides additional details for the construction of \Dname. 
In the main paper, we use the term \textit{clustering perspective} to denote a natural-language criterion for grouping audio recordings. 
In some data-generation prompts, we use the internal term \textit{dimension}; each retained dimension is converted into a clustering perspective in the benchmark.

\subsection{Shared Construction Pipeline}
\label{app:shared_pipeline}

All corpora follow the same high-level pipeline: perspective induction, perspective refinement, transcript construction, transcript validation, and audio synthesis.

\paragraph{Perspective Induction}
For each source corpus, we sample representative examples and prompt an LLM to propose candidate clustering perspectives. 
Each candidate's perspective contains a short description and a set of category labels. 
The goal is to induce perspectives that reflect meaningful reasoning factors in the corpus, such as communicative intent, interaction pattern, procedural structure, discourse function, risk type, or consequence, rather than superficial lexical overlap. 

\begin{promptbox}[Perspective generation prompt]
You are an expert in taxonomy design. Given (i) a description of a dataset and (ii) example text entries from it, your task is to propose meaningful clustering dimensions that capture distinct reasoning-based perspectives.
For each clustering dimension:
1. Provide a concise description of what the dimension represents.
2. List the possible cluster labels under this dimension.
3. Justify why each cluster is distinct, highlighting reasoning factors such as cause, intent, context, or outcome.
Requirements:
    - Propose 5--8 clustering dimensions.
    - For each dimension, propose 4--8 mutually exclusive cluster labels.
    - Clusters should be collectively as exhaustive as possible.
    - Avoid vague labels such as "Other", "Misc", or "General".
For each dimension, provide:
    - id
    - name
    - description
    - categories
For each category, provide:
    - id
    - label
    - description (what it is + why it's distinct)
Output STRICTLY as JSON with key "dimensions".
(i) Dataset description: {DATASET_DESCRIPTION}
(ii) Example text entries:
{EXAMPLES}
Please generate the clustering dimensions and taxonomies.
The JSON schema must be:
{
  "dimensions": [
    {
      "id": "dim_01_xxx",
      "name": "...",
      "description": "...",
      "categories": [
        {
          "id": "cat_01_xxx",
          "label": "...",
          "description": "..."
        }
      ]
    }
  ]
}
  \end{promptbox}

\paragraph{Perspective Refinement}
We refine each proposed perspective to ensure that its categories are mutually exclusive, semantically interpretable, and comparable in granularity. 
We remove perspectives with vague category boundaries, highly imbalanced categories, insufficient instance support, or categories that can be solved through shallow lexical cues. 
For each retained perspective, we generate a neutral clustering instruction that describes the grouping criterion without revealing the underlying category names.

\begin{promptbox}[Perspective refinement prompt]
You are a taxonomy refinement expert. Your goal is to revise the given taxonomy so that it is clear, consistent, and practically usable. Follow the requirements below:
Requirements:
    - Category name: Max {NAME_MAX_WORDS} words; concise, specific, and informative.
    - Description: Max {DESCRIPTION_MAX_WORDS} words; clearly explain what distinguishes this category.
    - Dimensions: Avoid near-duplicate dimensions that rely on the same primary signal. However, moderate correlation between dimensions is acceptable when conceptually justified.
    - Mutual exclusivity: Categories must not overlap or contradict each other. Within each dimension, categories should be defined so that most cases naturally fit ONLY ONE category.
    - Collective exhaustiveness: Categories together should cover all possible intents in the given context.
    - Granularity: All categories must be defined at the same level of specificity.
    - No vague labels: Avoid terms like "Other," "General," or "Miscellaneous."
    - Each category must be assignable using signals observable in the text or reasoning from the text.
[Dataset Context]
{DATASET_DESCRIPTION}
[Original Taxonomy]
{TAXONOMY_DRAFT}
Task:
1. Review the existing taxonomy and suggest improvements (e.g., renaming, merging, splitting, or adding new categories).
2. Ensure each category has a clear reasoning justification and aligns with the data context.
3. If categories are missing, add enough to make the taxonomy collectively exhaustive.
4. Based on the refined taxonomy, check the instruction and provide a neutral clustering instruction that accurately reflects the categorization principle without revealing category labels. Do not include or hint at the actual category names in the instruction.
Instruction Style Requirements:
    - For each dimension, generate one neutral clustering instruction.
    - Each instruction MUST start with "Cluster" or "Group"
    - Keep each instruction within 25 words.
    - Do NOT mention the taxonomy, dimensions, or categories explicitly. Do NOT include or hint at any category names.
    - Emphasize selecting the single most dominant factor in the case.
Output Format:
Return a JSON object with:
1. "refined_dimensions": The updated taxonomy list (same structure as input dimensions: id, name, description, categories with id, label, description).
2. "clustering_instructions": For each dimension, a neutral instruction (list or dict keyed by dimension id).
3. "change_log": A brief explanation of what was merged/split/added and why.
\end{promptbox}

\paragraph{Transcript Generation}
The transcript-generation procedure depends on the source format. 
For short intent queries, we expand the input into realistic spoken scripts. 
For long-form formal documents, we first compress the text into an evidence-preserving summary and then rewrite it into a spoken script. 
For dialogue-native corpora, we reuse the original dialogue structure. 
Across all corpora, the transcript is required to preserve the evidence needed for the assigned perspective labels. 

\paragraph{Transcript Validation}
We validate each generated transcript by asking independently prompted classifiers to assign it to a category under the corresponding perspective. 
An instance is retained only when the predicted label agrees with the intended label. 
This filtering step reduces label drift introduced by expansion, compression, or spoken-script rewriting. 
We also filter transcripts that explicitly mention category names, perspective names, or near-verbatim instruction phrases, preventing models from solving the task through lexical shortcuts.

\paragraph{Audio Synthesis and Paralinguistic Injection}
Validated transcripts are converted into speech using a TTS system (i.e., Qwen3-TTS). 
During synthesis, we inject controlled paralinguistic attributes, including emotion, speaker identity, speaker count, and background acoustic condition. 
For monologues, we synthesize a single speaker; for dialogues, we synthesize multiple speakers and concatenate turns according to the dialogue structure. 
For semantic perspectives, paralinguistic attributes are randomized and approximately balanced across categories. 
For paralinguistic perspectives, the corresponding acoustic attribute is used as the target clustering criterion, while other attributes are randomized to reduce confounding.

\subsection{Banking77}
\label{app:banking77}
Banking77~\cite{casanueva2020efficient} consists of short customer-service intent queries. 
Because the original queries are too short to serve as natural spoken recordings, we first classify each query under the refined perspectives and then expand classifiable queries into spoken scripts. 
The expanded scripts may be monologues, such as a customer describing a banking issue, or short dialogues, such as an exchange between a customer and an agent. 
The expansion prompt is constrained to preserve the assigned perspective labels while making the script realistic and conversational.

\begin{promptbox}[Script generation prompt for Banking77]
You are a specialized Banking AI with two distinct modes: a precise Analyst and a creative Scriptwriter.
### Mode 1: Precise Analyst
Analyze the input sentence against the provided taxonomy.
- USE CATEGORY NAME ONLY for classification.
- CRITICAL: If the sentence cannot be clearly mapped to a category in ANY of the dimensions, set that dimension to `None`.
- If ANY dimension is `None`, the `speech_script` field MUST be `None`.
[Dataset Context]
{DATASET_DESCRIPTION}
[Taxonomy]
{TAXONOMY}
### Mode 2: Creative Scriptwriter
Expand the content into a 150-200 word TTS script.
Goal: Create a diverse, realistic banking scenario.
Ambiguity: The script must be high-fidelity to the classification results. Avoid any content that would make the categories hard to distinguish.
Diversity Encouragement:
    - Speaker Choice: Feel free to use a Monologue (e.g., a customer's thought process, a voicemail) or a Dialogue (e.g., customer vs. agent, two friends discussing a bank issue). Let the context decide.
    - Dialogue Format:
        - If 2 speakers: Use "Speaker A: ..." and "Speaker B: ..." on new lines.
        - If 1 speaker: Use "Narrator: ..."
    - Narrative Flow: You are NOT limited to an FAQ format. You may weave the original concern into a story, a phone call, or a help-desk interaction.
    - Naturalism: Use spoken-language markers (pauses, "uhm", "right", "okay"). The script should sound like it's happening in real life.
    - Tone: Match the tone to the `Customer Harm and Urgency` dimension (e.g., calm for info, tense for security risks).
  \end{promptbox}

After expansion, we validate each script by re-classifying the generated transcript under the same perspective. 
Scripts are retained only if their transcript-level labels match the original query-level labels. 
The validated scripts are then synthesized into speech with sampled emotion, speaker identity, and background acoustic conditions. 
Speaker count is determined by the script format: monologues are synthesized with one speaker, while dialogues use multiple speakers.
\begin{promptbox}[Script validation prompt for Banking77]
You are a specialized Banking AI classifier. Your task is to analyze the provided speech script and classify it based on the taxonomy dimensions.
Classification Rules:
    - For each dimension in the taxonomy, determine which category the script belongs to.
    - Use ONLY the CATEGORY NAME from the provided taxonomy.
    - If the script cannot be clearly mapped to a category in ANY dimension, set that dimension to "None".
    - Crucial: You must also include the "original_sentence" provided in the input in your final JSON output without any changes.
[Dataset Context]
{DATASET_DESCRIPTION}
[Taxonomy]
{TAXONOMY}
Return ONLY a JSON object containing the "classification_results".
  \end{promptbox}

\subsection{ECHR}
\label{app:echr}

ECHR~\cite{poudyal2020echr} contains long-form legal case documents. 
We first sample case documents and induce legal-case perspectives from representative examples. 
The retained perspectives capture substantive legal and factual reasoning factors, such as the interest at stake, the type of state action, the procedural context, the affected population, or the form of harm.

Because the original documents are too long and formal for direct speech synthesis, we compress each accepted case into a shorter evidence-preserving summary. 
The compression is label-aware: the summary must retain the information required to recover all accepted perspective labels. 
We then validate the compressed summary by re-classifying it under the same perspectives. 
Only summaries whose labels remain recoverable are rewritten into spoken scripts.

The final spoken scripts are synthesized with one to four speakers, sampled emotion, and background acoustic conditions. 
Paralinguistic attributes are assigned in an approximately balanced way within semantic strata to avoid creating shortcuts between acoustic factors and semantic labels.

\subsection{S\&P 500}
\label{app:sp500}

The S\&P 500 corpus~\cite{jlohding_sp500_edgar_10k} consists of excerpts from SEC Form 10-K annual reports. 
The source texts cover company properties, legal proceedings, market information, dividends, and related stockholder matters. 
We induce clustering perspectives that reflect company-level patterns, such as asset ownership, geographic distribution, operational structure, litigation exposure, regulatory risk, shareholder payout policy, and capital-market behavior.

As with ECHR, the original disclosures are long and formal. 
We therefore compress each accepted text into a concise evidence-preserving summary and validate whether the accepted labels remain recoverable after compression. 
Validated summaries are then rewritten into spoken scripts. 
The synthesis stage injects emotion, speaker count, speaker identity, and background acoustic condition, while randomizing non-target paralinguistic attributes to reduce spurious correlations.

\subsection{MultiWOZ}
\label{app:multiwoz}

MultiWOZ~\cite{zang2020multiwoz} is already dialogue-native, so it does not require expansion from short queries or compression from long documents. 
We sample dialogues and induce perspectives that capture task-oriented interaction patterns, user goals, constraint evolution, request structure, and dialogue-level behavior. 
Each dialogue is labeled under the refined perspectives by independently prompted classifiers, and we retain only dialogues with consistent label assignments.

The original dialogue text is reused as the transcript. 
During TTS, each speaker role is assigned a distinct voice, and dialogue turns are synthesized in order. 
We additionally inject emotion and background acoustic variation. 
This preserves the multi-turn structure of MultiWOZ while converting it into a speech-based benchmark instance.

Tab.~\ref{tab:dataset_pipeline_summary} summarizes the corpus-specific transcript construction, validation, and TTS paths used in our synthesis pipeline.

\begin{promptbox}[Dataset description for taxonomy generation]
ECHR: This dataset contains English-language case documents from the European Court of Human Rights (ECtHR). Each entry is a legal case narrative written in formal legal style, describing factual background, procedural history, evidence, and legal context. The text often includes details about actors (individuals, authorities, courts), actions (detention, investigations, speech restrictions, searches, employment disputes, etc.), timelines, and outcomes in domestic proceedings. The goal is to discover reasoning-based clustering perspectives for grouping cases. Dimensions should reflect substantive legal and factual reasoning (e.g., the right or interest at stake, the type of state action, procedural context, affected population, type of harm, or dispute structure), rather than superficial keywords or document length.
S&P 500: This dataset contains excerpts from SEC Form 10-K annual reports of S&P 500 companies. Each entry is a section from a 10-K filing, written in formal financial/legal style. The data includes three item types:
    - Properties: Descriptions of company properties, facilities, real estate holdings, and physical assets.
    - Legal Proceedings: Disclosures of litigation, regulatory proceedings, and legal risks.
    - Market for Registrar's Common Equity: Information on stock performance, dividends, equity markets, and related stockholder matters.
The goal is to discover clustering perspectives for grouping these S&P 500 companies. Focus on company-level patterns reflected in the disclosures, such as:
    - asset ownership and utilization
    - geographic distribution of operations
    - industry-specific asset composition
    - exposure to litigation or regulatory risk
    - shareholder payout policies
    - capital structure or equity instruments
Banking77: This is a single-domain intent mining dataset consisting of user queries related to banking services. The dataset is characterized by subtle semantic differences between intents, where similar surface expressions may correspond to different underlying user goals. Focus on distinguishing fine-grained intent differences and capturing the exact user goal expressed in each query.
MultiWOZ: This is a multi-domain task-oriented dialogue dataset consisting of human-human conversations between a user and a system. Each dialogue spans one or more domains and involves completing tasks. The user's intent evolves across turns and may include constraints and requests. Focus on identifying the user's underlying intent at each turn and how it evolves across the dialogue, rather than treating each utterance independently.
\end{promptbox}

\begin{table*}[t]
\centering
\small
\resizebox{\linewidth}{!}{
\begin{tabular}{p{2.2cm}p{4.2cm}p{4.7cm}p{4.7cm}}
\toprule\toprule
Corpus & Transcript Generation & Validation & TTS Path \\
\midrule
Banking77 &
Classify each short intent query under the refined taxonomy, then expand each classifiable query into a natural spoken transcript (monologue or short dialogue). &
Re-classify the expanded transcript under the same taxonomy and retain it only if all transcript-level labels exactly match the original query-level assignments. &
Synthesize validated transcripts with sampled emotion and speaker identity; use one speaker for monologues and multiple speakers for dialogues; augment the waveform with background-noise conditions. \\
\midrule
ECHR &
Assign taxonomy labels to long legal case texts, compress each accepted case into an evidence-preserving summary, and rewrite the summary into a spoken script. &
Use dual-model agreement for source-text labeling, then re-classify the compressed text and retain it only if all previously accepted dimensions remain recoverable after compression. &
Synthesize accepted spoken scripts with one to four speakers, sampled emotion, and background-noise conditions, with approximate balancing of paralinguistic factors within semantic strata. \\
\midrule
S\&P 500 &
Assign taxonomy labels to 10-K disclosure segments, compress each accepted document into an evidence-preserving summary, and rewrite the summary into a spoken script. &
Use dual-model agreement for source-text labeling, then re-classify the compressed text and retain it only if all accepted dimensions are preserved after compression. &
Synthesize accepted spoken scripts with sampled emotion, speaker count, and background-noise condition, again using a balanced assignment scheme within semantic strata. \\
\midrule
MultiWOZ &
Directly reuse sampled human-written dialogues as transcripts after assigning taxonomy labels. &
Retain only dialogues for which two independent classifiers produce identical label assignments across all taxonomy dimensions. &
Synthesize the validated dialogue turn by turn with distinct speakers for different roles, while injecting sampled emotion and background-noise variation. \\
\bottomrule\bottomrule
\end{tabular}
}
\caption{Corpus-specific transcript generation, validation, and TTS paths. Banking77 uses expansion from short intent queries; ECHR and S\&P 500 use compression followed by spoken rewriting; MultiWOZ directly reuses native dialogues.}
\label{tab:dataset_pipeline_summary}
\end{table*}

\subsection{Quality Control}
\label{app:quality_control}

We apply several quality-control steps to make the benchmark reliable and to reduce shortcut learning.

First, each retained perspective must define categories that are mutually exclusive, interpretable, and sufficiently supported by examples. 
Perspectives with ambiguous decision boundaries, severe class imbalance, or categories that rely on superficial lexical cues are removed.

Second, transcript construction is validated by label-preservation checks. 
After expansion, compression, or rewriting, each transcript is re-classified under the corresponding perspective. 
Only transcripts whose labels can be recovered are retained.

Third, we filter explicit lexical leakage. 
Generated transcripts must not directly contain perspective names, category names, or near-verbatim instruction phrases. 
This prevents models from relying on trivial string matching.

Fourth, paralinguistic attributes are controlled during synthesis. 
For semantic perspectives, acoustic attributes are randomized and approximately balanced across categories. 
For paralinguistic perspectives, the target acoustic attribute defines the label, while other attributes are randomized. 
This reduces unintended correlations between semantic labels and acoustic conditions.

Tab.~\ref{tab:dataset_perspectives} shows the representative perspectives of each adopted corpus.

\subsection{TTS system}
\label{app:tts}
We used Qwen3-TTS CustomVoice to generate audio from the transcripts. 
During the synthesis, the model accepts, per utterance, the input text, a fixed language argument (``English'' throughout the evaluation pipelines), target speakers, and a natural-language instruction that conditions the prosodic style. 

\paragraph{Speaker Voices and Multi-Speaker Rendering}
Speaker identity is drawn from a fixed inventory of nine CustomVoice speakers: Vivian, Serena, Uncle\_Fu, Dylan, Eric, Ryan, Aiden, Ono\_Anna, and Sohee. 
During generation, each item is assigned a speaker count in \{1, 2, 3, 4\}, and a corresponding set of distinct speaker identities sampled from this pool, up to the size of the inventory. 
Single-speaker items use the label ``narrative'', whereas multi-speaker items rotate through speakerA, speakerB, speakerC, and speakerD in strict cyclic order enforced by the generation prompt.

At synthesis time these abstract labels are bound to concrete Qwen speakers. 
For monologues, the full narrative text is rendered in a single call using the first assigned speaker identity. 
For conversations, the unique speaker labels are extracted in order of first appearance and mapped positionally onto the assigned speaker identities. 
Each dialogue segment is then synthesized independently with its bound speaker, and the resulting per-segment waveforms are concatenated with a fixed 0.3-second silence inserted between turns to demarcate speaker changes and impart conversational pacing.

\paragraph{Emotion Synthesis}
Emotional coloring is realized through instruction conditioning. 
Each item is assigned one of five categorical emotions: happiness, sadness, surprise, anger, or neutral. 
At synthesis time, the assigned emotion is mapped to a short natural-language directive that is passed as the instruct argument to the synthesizer (for example, ``Speak in a happy, warm tone.'' for happiness).

\paragraph{Background-Noise Injection}
Following clean synthesis, each waveform is optionally degraded with an environmental background track to emulate realistic recording conditions. 
The noise taxonomy comprises four categories: clean (no degradation), indoor ambience, crowd noise, and traffic noise.
For non-clean items, a noise file is drawn at random from the category-specific pool indexed over the configured noise directories. 
The chosen noise is converted to mono, resampled to the speech sample rate, and either randomly cropped or tiled with a random offset to match the speech length.

Mixing is performed at a controlled signal-to-noise ratio (SNR). 
The noise segment is first RMS-normalized, and its gain is then set so that the mixture attains the target SNR, following the relation 
\begin{equation}
    desired\_noise\_rms = \frac{speech\_rms}{ 10^{\frac{snr\_db}{20}}}.
\end{equation}

To prevent clipping, the summed signal is peak-limited: if the maximum absolute amplitude exceeds 0.99, the mixture is rescaled by 0.99 / peak. 
If a per-item $snr\_db$ is absent, an SNR is drawn uniformly from a configurable range (default 10–20 dB). 

\paragraph{Sampling and Stratified Assignment Rules}
The assignment of speaker count, emotion, and background-noise category is governed by a stratified, approximately uniform sampling scheme designed to mitigate joint skew across task types and taxonomy categories. 
Only items marked ``accepted'' with non-empty compressed text are eligible. 
Items are partitioned into strata defined by a stratum key.

Within each stratum of size $m$, a balanced routine allocates the $m$ positions as evenly as possible over the candidate values of each attribute independently: it computes base, $rem = \textbf{divmod}(m, k)$ for $k$ categories, assigns base occurrences to every category plus one extra to the first rem categories, and then shuffles the resulting multiset. 
The three per-attribute multisets—speaker counts, emotions, and noise categories—are zipped into triples, which are themselves shuffled to decorrelate the marginal assignments. 
Consequently, each attribute is close to uniformly distributed within every taxonomy stratum, rather than merely globally. 
An SNR value is drawn ($\textbf{uniform}(10.0, 20.0)$, rounded to two decimals) whenever the assigned noise category is not clean, and is left null otherwise. 

\begin{table*}[t]
\centering
\scriptsize
\setlength{\tabcolsep}{4pt}
\renewcommand{\arraystretch}{1.05}
\begin{tabular}{p{1.5cm} p{6.25cm} p{6.25cm}}
\toprule
\toprule
\textbf{Corpus} & \textbf{Held-in perspectives} & \textbf{Held-out perspectives} \\
\midrule

\textbf{Banking77}
&
\begin{itemize}[leftmargin=1em,nosep,itemsep=0pt,topsep=0pt,parsep=0pt]
    \item Cluster requests by the customer’s primary intended outcome
    \item Cluster requests by the main banking problem or request category
    \item Cluster requests by the dominant urgency, severity, or risk level
    \item Cluster clips by the dominant emotional tone in speech
    \item Cluster clips by the exact number of distinct speakers
    \item Cluster clips by the real-world background-noise scene
\end{itemize}
&
\begin{itemize}[leftmargin=1em,nosep,itemsep=0pt,topsep=0pt,parsep=0pt]
    \item Cluster requests by the banking capability or product domain most directly involved
    \item Cluster requests by the primary route through which the issue would be resolved
    \item Cluster clips by the dominant speaker-gender pattern
\end{itemize}
\\

\textbf{ECHR}
&
\begin{itemize}[leftmargin=1em,nosep,itemsep=0pt,topsep=0pt,parsep=0pt]
    \item Cluster cases by the dominant State act or omission producing the alleged violation
    \item Cluster cases by the procedural or situational context of the dispute
    \item Cluster cases by the structure of the dispute and party configuration
    \item Cluster clips by the dominant emotional tone in speech
    \item Cluster clips by the exact number of distinct speakers
    \item Cluster clips by the real-world background-noise scene
\end{itemize}
&
\begin{itemize}[leftmargin=1em,nosep,itemsep=0pt,topsep=0pt,parsep=0pt]
    \item Cluster cases by the principal protected interest implicated
    \item Cluster cases by the main kind of harm or outcome alleged
    \item Cluster cases by the most salient applicant status affecting power imbalance or risk
    \item Cluster clips by the dominant speaker-gender pattern
\end{itemize}
\\

\textbf{S\&P 500}
&
\begin{itemize}[leftmargin=1em,nosep,itemsep=0pt,topsep=0pt,parsep=0pt]
    \item Cluster firms by the primary way key properties are controlled
    \item Cluster firms by the geographic breadth and concentration of properties
    \item Cluster firms by the dominant type of legal matter discussed
    \item Cluster firms by the dominant signal about legal-matter status and potential financial impact
    \item Cluster clips by the dominant emotional tone in speech
    \item Cluster clips by the exact number of distinct speakers
    \item Cluster clips by the real-world background-noise scene
\end{itemize}
&
\begin{itemize}[leftmargin=1em,nosep,itemsep=0pt,topsep=0pt,parsep=0pt]
    \item Cluster disclosures by the dominant type of physical properties central to operations
    \item Cluster disclosures by the dominant approach to shareholder return
    \item Cluster disclosures by the dominant approach to repurchases or equity structure
    \item Cluster clips by the dominant speaker-gender pattern
\end{itemize}
\\

\textbf{MultiWOZ}
&
\begin{itemize}[leftmargin=1em,nosep,itemsep=0pt,topsep=0pt,parsep=0pt]
    \item Cluster dialogues by the presence and source of explicit repair turns
    \item Cluster dialogues by how users provide, accumulate, revise, or confirm constraints
    \item Cluster dialogues by the balance between information-seeking and transaction-execution content
    \item Cluster dialogues by whether the task succeeds directly, partially succeeds, fails, or is recovered through fallback
    \item Cluster clips by the dominant emotional tone in speech
    \item Cluster clips by the exact number of distinct speakers
    \item Cluster clips by the real-world background-noise scene
\end{itemize}
&
\begin{itemize}[leftmargin=1em,nosep,itemsep=0pt,topsep=0pt,parsep=0pt]
    \item Cluster dialogues by how much earlier information must be remembered and reused later
    \item Cluster dialogues by who drives the interaction and how strongly negotiation shapes the exchange
    \item Cluster dialogues by whether the interaction stays within a narrow goal or couples multiple domains or subgoals
    \item Cluster clips by the dominant speaker-gender pattern
\end{itemize}
\\

\bottomrule
\bottomrule
\end{tabular}
\caption{Corpus-specific clustering perspectives.}
\label{tab:dataset_perspectives}
\end{table*}

\section{Benchmark Split Construction}
\label{app:split_construction}

\subsection{Held-in and Held-out Perspectives}

We partition perspectives rather than only individual audio recordings. 
Held-in perspectives are available during training, including their natural-language instructions and category taxonomies. 
Held-out perspectives are never used during training and are reserved for evaluating whether a model can adapt to novel clustering criteria. 
This design prevents evaluation from only measuring memorization of familiar label structures.

\subsection{Training Instance Construction}

For each held-in perspective, we split audio recordings within every category into a training pool and an evaluation pool. 
Training instances are generated by first sampling a subset of categories and then sampling multiple audio recordings from each selected category. 
This produces clustering instances with varying category combinations and cluster sizes. 
The model is not given the number of clusters; it must infer the number of clusters from the input audio and the natural-language perspective.

\subsection{Evaluation Levels}

The evaluation set is divided into three levels.

\paragraph{$\mathrm{\textbf{L}}_0$: seen perspectives and seen audio.}
$\mathrm{\textbf{L}}_0$ is constructed by re-sampling category and audio combinations from the training pool of held-in perspectives, while ensuring that the resulting clustering instances do not appear during training. 
This level evaluates recombination robustness: the model sees the same perspectives and audio recordings during training, but must solve new clustering combinations.

\paragraph{$\mathrm{\textbf{L}}_1$: seen perspectives and unseen audio.}
$\mathrm{\textbf{L}}_1$ is constructed from the evaluation pool of held-in perspectives. 
The clustering perspectives and their taxonomies are familiar, but none of the audio recordings appear in the training instances. 
This level evaluates whether the model can generalize to new recordings under familiar clustering criteria.

\paragraph{$\mathrm{\textbf{L}}_2$: unseen perspectives and unseen audio.}
$\mathrm{\textbf{L}}_2$ is drawn exclusively from held-out perspectives. 
The model has not observed the perspective instructions, category taxonomies, or audio recordings during training. 
This level evaluates cross-perspective generalization, namely whether the model can adapt to new natural-language clustering criteria over unseen audio collections.

\subsection{Episode Construction and Benchmark Statistics}
\label{app:benchmark_stats}

\paragraph{Episode formulation.}
Each benchmark instance is formulated as an episode consisting of a
natural-language clustering perspective \(p\) and \(n\) indexed audio
clips, where \(4 \leq n \leq 10\). The model input is represented as
\[
    \left(
    p,\,
    \{\lvert i\rvert \langle\text{audio}_{i}\rangle\}_{i=1}^{n}
    \right),
\]
and the target output is a partition
\(\mathcal{C}=\{C_1,\ldots,C_K\}\) over the \(n\) clips.
The semantic category names used to construct the partition are hidden
from the model. Moreover, the gold number of clusters \(K\) is not
provided. The model must therefore jointly infer the partition
cardinality and assign every input clip to exactly one cluster. This
formulation evaluates both perspective-conditioned semantic
understanding and variable-cardinality clustering.

\paragraph{Controlled episode sampling.}
We construct episodes using a two-stage sampling procedure. First,
category-specific audio pools are created for each clustering
perspective. For held-in perspectives, the recordings are divided into
training and evaluation pools, whereas held-out perspectives are
reserved exclusively for evaluating $\mathrm{\textbf{L}}_2$ generalization. This separation
ensures that generalization to unseen perspectives is evaluated
independently from the construction of the SFT training episodes.

Second, episodes are generated through quota-aware perspective
selection and adaptive balancing over the feasible values of \(K\).
Trivial single-cluster episodes are explicitly down-weighted, while
categories with lower sampling coverage are assigned higher priority.
After sampling the constituent categories, the total number of clips is
drawn conditionally on \(K\): episodes with smaller \(K\) preferentially
contain \(4\)--\(6\) clips, whereas episodes with larger \(K\) may
contain up to \(10\) clips. Candidates with invalid clip counts or
duplicate episode signatures are rejected and resampled. This procedure
balances category coverage while retaining natural variation in both
episode size and partition cardinality.

\begin{table*}[t]
    \centering
    \small
    \setlength{\tabcolsep}{5.5pt}
    \caption{
        Episode-level statistics of the SFT training set.
        Entries for the final three columns are reported as
        mean (95th percentile, maximum).
    }
    \label{tab:sft-episode-statistics}
    \begin{tabular}{lrrrrr}
        \toprule
        \toprule
        Source
        & Episodes
        & Unique audios
        & Clips / episode
        & Gold \(K\)
        & Duration (s) \\
        \midrule
        All SFT train
        & 1,876 & 2,008
        & 5.64 (7, 10)
        & 3.37 (6, 9)
        & 373.5 (524.6, 539.9) \\
        ECHR
        & 539 & 501
        & 5.51 (7, 7)
        & 3.37 (6, 7)
        & 452.8 (531.4, 539.9) \\
        S\&P 500
        & 571 & 455
        & 5.69 (7, 9)
        & 3.45 (6, 9)
        & 411.4 (525.2, 539.2) \\
        Banking77
        & 523 & 618
        & 5.88 (8, 10)
        & 3.66 (7, 9)
        & 306.3 (431.1, 538.1) \\
        MultiWOZ
        & 243 & 434
        & 5.33 (7, 7)
        & 2.56 (4, 5)
        & 253.0 (372.3, 445.3) \\
        \bottomrule
        \bottomrule
    \end{tabular}
\end{table*}

\paragraph{Episode-level statistics.}
As summarized in Tab.~\ref{tab:sft-episode-statistics}, the SFT
training set contains 1876 episodes constructed from 2008 unique audio recordings. Based on the mean episode size, these episodes comprise approximately 10.6K clip occurrences, corresponding to an average reuse factor of approximately \(5.3\times\). 
Such recombination allows the same recording to participate in different episode configurations and prevents the training set from reducing to a fixed collection of partitions.

\begin{figure}[tb]
    \centering
    % Use 0.95\columnwidth for a single-column figure.
    % For a wider single-column document, it can be changed to
    % 0.69\textwidth.
    \resizebox{\columnwidth}{!}{%
        \begin{tikzpicture}
            \begin{polaraxis}[
                % Equal width and height ensure a circular plot.
                width=8.5cm,
                height=8.5cm,
                scale only axis,
                axis equal image,
                clip=false,
                ymin=0,
                ymax=100,
                ytick={0,25,50,75,100},
                xtick={0,72,144,216,288},
                xticklabels={
                    {\quad \(\boldsymbol{K>1}\)\\ \quad episodes},
                    {Singleton\\clusters},
                    {Size-2\\clusters},
                    {Size-\(\boldsymbol{\geq 3}\)\\clusters},
                    {Episodes with\\non-singletons}
                },
                x tick label style={
                    font=\scriptsize,
                    align=center
                },
                y tick label style={
                    font=\scriptsize
                },
                grid=both,
                major grid style={
                    draw=gray!50,
                    line width=0.4pt
                },
                axis line style={
                    draw=black,
                    line width=0.6pt
                },
                legend style={
                    at={(0.5,-0.1)},
                    anchor=north,
                    legend columns=2,
                    draw=none,
                    font=\scriptsize,
                    column sep=8pt
                }
            ]

                % ECHR
                \addplot+[
                    thick,
                    mark=*,
                    mark size=2.2pt,
                    blue!75!black
                ] coordinates {
                    (0,95.2)
                    (72,54.7)
                    (144,29.1)
                    (216,16.2)
                    (288,91.7)
                    (360,95.2)
                };
                \addlegendentry{ECHR}

                % S&P 500
                \addplot+[
                    thick,
                    mark=square*,
                    mark size=2.2pt,
                    orange!85!black
                ] coordinates {
                    (0,96.5)
                    (72,53.2)
                    (144,30.5)
                    (216,16.3)
                    (288,94.2)
                    (360,96.5)
                };
                \addlegendentry{S\&P 500}

                % Banking77
                \addplot+[
                    thick,
                    mark=triangle*,
                    mark size=2.4pt,
                    green!55!black
                ] coordinates {
                    (0,95.6)
                    (72,55.5)
                    (144,31.2)
                    (216,13.3)
                    (288,93.9)
                    (360,95.6)
                };
                \addlegendentry{Banking77}

                % MultiWOZ
                \addplot+[
                    thick,
                    mark=diamond*,
                    mark size=2.5pt,
                    violet
                ] coordinates {
                    (0,92.2)
                    (72,32.1)
                    (144,38.0)
                    (216,29.9)
                    (288,100.0)
                    (360,92.2)
                };
                \addlegendentry{MultiWOZ}

            \end{polaraxis}
        \end{tikzpicture}%
    }

    \caption{
        Source-specific episode and cluster profiles in the SFT
        training set. All axes report percentages. The first axis
        represents the percentage of multi-cluster episodes, i.e.,
        \(100\% - \Pr(K=1)\). The three cluster-size axes form a
        compositional distribution and therefore sum to \(100\%\)
        for each source.
    }
    \label{fig:sft-cluster-profile}
\end{figure}
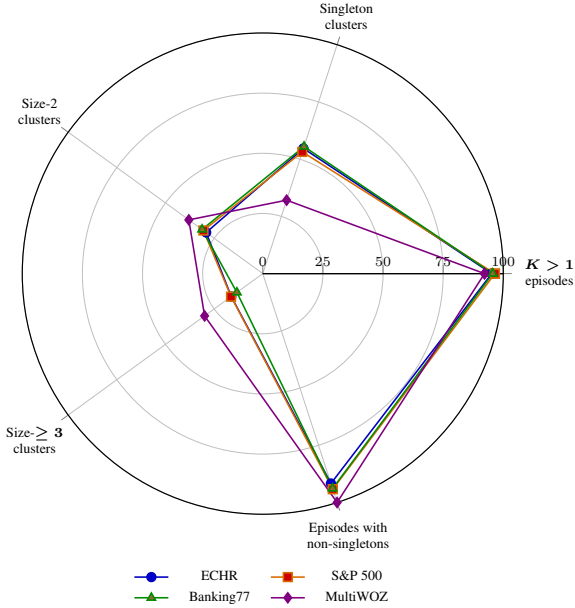

The training episodes contain 5.64 clips on average, with a 95th percentile of 7 clips, while the corresponding gold partitions contain 3.37 clusters on average and up to 9 clusters. 
The ratio between the aggregate mean episode size and mean \(K\) is approximately 1.67 clips per cluster, indicating that the benchmark predominantly evaluates compact, fine-grained partitions rather than a small number of large clusters. 
The average episode duration is 373.5 seconds (\(6.2\) minutes), and the maximum duration is approximately 9 minutes.
Thus, the task additionally requires reasoning over relatively long multi-audio contexts.

The individual sources exhibit complementary forms of complexity.
Banking77 produces the largest episodes and the highest average partition cardinality, with 5.88 clips and 3.66 clusters per episode.
ECHR instead contributes the longest temporal contexts, averaging 452.8 seconds despite containing fewer clips than Banking77.
S\&P~500 lies between these two regimes. 
MultiWOZ produces shorter episodes with a substantially smaller mean \(K\) of 2.56, yielding denser clusters with approximately 2.08 clips per cluster. 
These differences prevent the training distribution from being dominated by a single notion of episode difficulty.

\paragraph{Cluster-size distribution.}
Across the complete training set, the 1876 episodes contain 6323 gold clusters. 
Singleton clusters account for 52.3\% of these clusters, whereas clusters of size two and size three or larger account for 31.1\% and 16.7\%, respectively. 
The prevalence of singleton clusters is expected because a typical episode distributes only five to seven clips across multiple semantic categories. 
Importantly, the cluster-level singleton rate should not be interpreted as evidence of episode-level degeneracy.
Only 4.7\% of the episodes have \(K=1\), while 94.1\% contain at least one non-singleton cluster. 
Consequently, the vast majority of episodes require the model to identify at least one positive within-cluster relation while simultaneously separating clips belonging to different categories.

Fig.~\ref{fig:sft-cluster-profile} visualizes the percentage-based cluster-structure statistics across the four sources. 
MultiWOZ exhibits a notably denser partition structure: only 32.1\% of its gold clusters are singletons, compared with 53.2--55.5\% for the other three sources, and 29.9\% contain at least three clips. 
In contrast, Banking77 has the largest average \(K\) but the smallest proportion of size-three-or-larger clusters, indicating that its higher cardinality primarily arises from a larger number of fine-grained categories rather than larger within-category groups.

\section{Additional Method Details}
\label{app:method_details}

This appendix provides additional details for the training procedure of \Mname. 
Appendix~\ref{app:rd_details} describes reasoning distillation, and Appendix~\ref{app:dpo_details} describes preference-pair construction and DPO training.

\subsection{Reasoning Distillation Details}
\label{app:rd_details}

\paragraph{Gold Partition Construction}
For each training instance, benchmark annotations provide a category label for every audio segment under the corresponding clustering perspective. 
We convert these labels into a gold partition over item indices: two segments are assigned to the same cluster if and only if they share the same category label. 
This representation removes dependence on category names and cluster order, allowing supervision and evaluation to operate directly on partition structure.

\paragraph{Teacher Trace Generation}
Given an input audio set $\mathcal{D}^{(i)}$, a clustering perspective $p^{(i)}$, and the gold partition $\mathcal{C}^{(i)*}$, we ask a teacher model to generate a concise reasoning trace that naturally leads to the gold partition. 
The teacher is instructed to identify the grouping principle, compare segments across the set, resolve confusable cases, infer the number of clusters, and end with a final clustering block that assigns every item exactly once.

This procedure can be viewed as gold-constrained rationale synthesis. 
The teacher is not asked to discover the answer from scratch; instead, it is conditioned on the correct partition and asked to synthesize a plausible reasoning path consistent with that partition. 
This improves supervision quality by separating answer discovery from explanation generation.

\paragraph{Linguistic and Paralinguistic Traces}
We construct reasoning traces from two complementary views. 
For linguistic perspectives, we transcribe each audio segment and provide the teacher with indexed text inputs. 
The teacher then generates a semantically grounded rationale based on the lexical content. 
For paralinguistic perspectives, we provide indexed audio clips directly to an audio-capable teacher model. 
The teacher is encouraged to ground its reasoning in audible evidence. 
When both views are available for an instance, we retain both traces as complementary supervision. 

\begin{promptbox}[Linguistic reasoning generation prompt]
You are a clustering assistant for audio-style clustering.
Given a clustering goal and a list of indexed items (|1|. ..., |2|. ..., |3|. ...),
produce a reasoning process that clusters them correctly.
The clustering uses 1-based indexing for all items.
Think process of the task:
First, go through the items and identify the most plausible grouping principle under
the clustering goal. Form an initial view of how many clusters there may be and what
distinguishes them.
Then compare items across the set, focusing on which items belong together, which
ones are easily confusable, and what distinctions matter most.
If needed, include one brief moment of uncertainty, revision, or self-check, but
only when it naturally helps resolve a genuine grouping decision.
Finally, state the final grouping so that every item belongs to exactly one cluster.
Clustering Goal:
{INSTRUCTION}
Items:
{ENUMERATED_TEXT}
The correct final clustering results are:
<answer>
{CORRECT_ANSWER}
</answer>
Now produce a high-quality think process that naturally leads to the correct clustering.
IMPORTANT!!!
1. Write strictly from a direct listening perspective.
2. Do NOT mention text, transcripts, reading, or simulated listening.
3. Do NOT use phrasing that implies prior knowledge of the final answer (for example: "looking at the correct answer", "we must match", "the real clusters are", "since the answer is given").
4. Do NOT perform explicit sanity checks against the provided answer block.
5. The correct clustering should emerge after comparison and reasoning, not be stated immediately.
6. The reasoning should not start from fixed category names or predefined labels.
7. Cross-item comparison is required.
8. Include uncertainty, backtracking, or self-verification only when it naturally arises; do not force it.
9. Avoid rigid template repetition or purely item-by-item labeling.
10. Do not describe every item individually in sequence unless absolutely necessary; prioritize group-level comparison.
11. Keep the reasoning very concise, with high information density (useful grouping detail, not filler); the full reasoning should be between 120 and 200 words.
12. Avoid repeated paraphrases, repeated pairwise comparisons, or long explanations that do not add new grouping insight.
13. Each paragraph should contribute either:
    a grouping hypothesis,
    a comparison that separates or merges items,
    or a final assignment decision.
14. End with a clearly recoverable final grouping that covers all items exactly once.
Output format:
- Start with: Think process:
- End with a short final clustering block in this style:
Final clustering:
- Cluster 1: [ ... ]
- Cluster 2: [ ... ]
...
- Do not include anything outside the reasoning process.
  \end{promptbox}

  \begin{promptbox}[Paralinguistic reasoning generation prompt]
You are a clustering assistant for audio clustering.
Given a clustering goal and indexed audio clips (|1|, |2|, |3|, ...), listen to
the clips and produce a reasoning process that clusters them correctly.
The clustering uses 1-based indexing for all items.
Think process of the task:
First, listen through the clips and identify the most plausible grouping principle under the clustering goal. Form an initial view of how many clusters there may be and what audible cues distinguish them.
Then compare clips across the set, focusing on which clips belong together, which ones are easily confusable, and what distinctions matter most.
If needed, include one brief moment of uncertainty, revision, or self-check, but only when it naturally helps resolve a genuine grouping decision.
Finally, state the final grouping so that every clip belongs to exactly one cluster.
Clustering Goal:
{INSTRUCTION}
There are {N_ITEMS} clips in total, indexed from 1 to {N_ITEMS}.
The correct final clustering results are:
<answer>
{CORRECT_ANSWER}
</answer>
Now produce a high-quality think process that naturally leads to the correct clustering.
IMPORTANT!!!
1. Ground the reasoning in audible evidence such as voice characteristics, prosody, timing, overlap, background sounds, acoustic scene cues, or other perceptual clues relevant to the task.
2. Do NOT rely on semantic content unless the task itself is explicitly semantic.
3. Do NOT use phrasing that implies prior knowledge of the final answer.
4. Do NOT perform explicit sanity checks against the provided answer block.
5. The correct clustering should emerge after comparison and reasoning, not be stated immediately.
6. The reasoning should not start from fixed category names or predefined labels.
7. Cross-clip comparison is required.
8. Include uncertainty, backtracking, or self-verification only when it naturally arises; do not force it.
9. Avoid rigid template repetition or purely clip-by-clip labeling.
10. Do not describe every clip individually in sequence unless absolutely necessary; prioritize group-level comparison.
11. Keep the reasoning very concise, with high information density (useful grouping detail, not filler); the full reasoning should be between 120 and 200 words.
12. Avoid repeated paraphrases, repeated pairwise comparisons, or long explanations that do not add new grouping insight.
13. Each paragraph should contribute either:
    a grouping hypothesis,
    a comparison that separates or merges clips,
    or a final assignment decision.
14. End with a clearly recoverable final grouping that covers all clips exactly once.
Output format:
    - Start with: Think process:
    - End with a short final clustering block in this style:
Final clustering:
    - Cluster 1: [ ... ]
    - Cluster 2: [ ... ]
...
- Do not include anything outside the reasoning process.
  \end{promptbox}

\paragraph{Trace Constraints}
We impose several constraints on teacher-generated traces. 
First, the reasoning must be comparison-driven rather than a sequence of independent item labels. 
Second, it should explain why some segments belong together and why others should be separated. 
Third, it must not mention that the gold answer is provided or imply that the reasoning is merely matching a known solution. 
Fourth, it should be concise and information-dense, avoiding rigid templates or repeated paraphrases. 
Finally, paralinguistic traces must be grounded in audible evidence rather than generic semantic descriptions.

\paragraph{Verification and Filtering}
We apply a two-stage filtering procedure before adding a trace to the RD dataset. 
First, we parse the final clustering block and compare the implied partition with the gold partition, ignoring cluster order and cluster names. 
If rule-based parsing is inconclusive, we use an independent verifier model to determine whether the final grouping matches the gold partition. 
Second, we use a judge model to evaluate the reasoning quality, including completeness, logical coherence, comparison quality, decision quality, conciseness, and modality consistency. 
Only traces that pass both the partition-consistency check and the quality judgment are retained.

\begin{promptbox}[Partition checking prompt]
You are a strict partition checker.
Your ONLY task is to decide whether the FINAL clustering assignment implied by the reasoning is exactly the same partition as the ground truth.
Rules:
1. Ignore narrative quality, fluency, confidence, or how convincing the explanation sounds.
2. Focus only on the final grouping the author commits to.
3. If the final grouping is ambiguous, incomplete, or not recoverable, output MISMATCH.
4. Cluster order does NOT matter.
5. Category names do NOT matter.
6. Only the partition over item indices matters.
Items are indexed 1..{n_items}, and each item must appear exactly once.
Ground truth (correct partition):
{correct_answer}
Reasoning chain to check:
{reasoning_chain}
Determine whether the final grouping implied by the reasoning matches the ground truth exactly as a partition over indices.
Output exactly:
VERDICT: MATCH or MISMATCH
REASON: one short sentence describing only the grouping comparison.
  \end{promptbox}

\begin{promptbox}[Reasoning quality evaluation prompt]
You are an expert AI judge tasked with evaluating the quality and completeness of clustering reasoning chains.
Evaluate the reasoning chain using the following criteria:
1. Completeness:
   Does the reasoning provide a full path from initial observations to grouping decisions and a final assignment?
2. Logical Coherence:
   Does the reasoning proceed in a consistent and understandable sequence, without major contradictions or unexplained jumps?
3. Comparison Quality:
   Does the reasoning compare items across the set, rather than merely describing or labeling each item independently?
4. Decision Quality:
   Does the reasoning actually make grouping decisions, including separating confusable items or merging similar ones for clear reasons?
5. Exploration Quality:
   If uncertainty or ambiguity naturally arises, is it handled briefly and usefully? Reject fake or padded uncertainty.
6. Modality Consistency:
   - For linguistic reasoning outputs, the reasoning should still read like direct listening-based reasoning and should not mention reading or transcripts.
   - For paralinguistic reasoning outputs, the reasoning should be grounded in audible evidence rather than a generic semantic paraphrase.
7. Final Assignment Presence:
   Does the reasoning clearly imply a final clustering that assigns every indexed
   item exactly once?
8. Conciseness and Information Density:
   The reasoning should be concise and information-dense. Reject if it contains significant redundancy, repeated comparisons, repeated paraphrasing, or long passages that add little new grouping information. Do NOT reject solely because it is somewhat long if it remains efficient and informative.
Reject if any of the following is true:
    - The chain is too short, incomplete, or abruptly cut off
    - There is no real cross-item comparison
    - The reasoning is mostly template-like or mostly item-by-item labeling
    - It explicitly refers to reading, transcripts, or matching the provided answer
    - It forces artificial uncertainty or artificial self-correction
    - The final implied clustering is missing, incomplete, or does not cover all items exactly once
    - The chain is overly verbose relative to the amount of actual grouping insight
Reasoning chain to evaluate:
{reasoning_chain}
Response format:
ASSESSMENT: ACCEPT or REJECT
REASON: one concise explanation
  \end{promptbox}
  
\paragraph{Training Target}
Each retained RD target contains a reasoning trace followed by a final clustering answer. 
The model is trained with the autoregressive loss:
\begin{equation}
\mathcal{L}_{\mathrm{RD}}(\theta)
=
-
\sum_{(x,y)\in \mathcal{D}_{\mathrm{RD}}}
\sum_{t=1}^{|y|}
\log p_\theta(y_t \mid x,y_{<t}).
\end{equation}
Although the RD target includes reasoning, the final answer is always serialized in a canonical clustering format so that downstream evaluation can recover the predicted partition.

\subsection{Preference Optimization Details}
\label{app:dpo_details}

\paragraph{Preference-pair Construction}
For each training input $x$, we construct a preference pair $(x,y^+,y^-)$. 
The chosen response $y^+$ is the canonical gold clustering answer derived from the benchmark annotation. 
The rejected response $y^-$ is sampled from the RD model. 
A sampled response is eligible only if it satisfies three conditions: it is parsable as a clustering answer, it assigns every item exactly once with no missing, duplicate, or unknown indices, and it produces a partition different from the gold partition.

This validity filter prevents DPO from being dominated by trivial formatting mistakes. 
Instead, the preference objective focuses on valid but incorrect clustering decisions.

\paragraph{Answer-token Margin}
To identify hard rejected answers, we compute the answer-token mean log-probability margin under the initial model $\pi_0$:
\begin{equation}
m(x,y^+,y^-)
=
\frac{\log \pi_0(y^+\mid x)}{|y^+|}
-
\frac{\log \pi_0(y^-\mid x)}{|y^-|}.
\end{equation}
A small or negative margin indicates that the initial model assigns comparable or higher likelihood to the rejected answer than to the gold answer. 
We therefore prioritize pairs with small margins, since they expose errors that the model is likely to make at inference time.

\paragraph{Clustering-quality Gap}
We also compute a clustering-quality gap
\begin{equation}
g = s(y^+) - s(y^-),
\end{equation}
where $s(\cdot)$ is a clustering quality score used for candidate filtering. 
Candidates with extremely small gaps are removed because they may correspond to nearly equivalent partitions or ambiguous cases. 
The remaining candidates are valid but meaningfully worse than the gold clustering.

\paragraph{Structural Error Categories}
To balance the DPO data across different types of clustering errors, each valid rejected answer is assigned to one structural error category. 
Let $K^+$ and $K^-$ denote the gold and predicted numbers of clusters. 
Let $R_{\mathrm{same}}$ be the recall of gold same-cluster pairs, and let $R_{\mathrm{diff}}$ be the recall of gold different-cluster pairs. 
We categorize rejected answers using the following deterministic rules:
\begin{equation}
\resizebox{\columnwidth}{!}{$
\displaystyle
e(y^{-}) =
\begin{cases}
\textsc{near-miss}, 
& 0.20 \leq g \leq 0.40 \ \land\ K^{-}=K^{+}, \\
\textsc{over-merge}, 
& K^{-}<K^{+} \ \lor\ (R_{\mathrm{same}}\geq 0.80 \land R_{\mathrm{diff}}<0.60), \\
\textsc{over-split}, 
& K^{-}>K^{+} \ \lor\ (R_{\mathrm{diff}}\geq 0.80 \land R_{\mathrm{same}}<0.60), \\
\textsc{k-wrong}, 
& K^{-}\neq K^{+}, \\
\textsc{wrong-assignment}, 
& \text{otherwise}.
\end{cases}
$}
\end{equation}

\textsc{Near-miss} cases have the correct number of clusters but imperfect assignments. 
\textsc{Over-merge} cases collapse distinct gold clusters, while \textsc{over-split} cases fragment a gold cluster into multiple predicted clusters. 
\textsc{K-wrong} captures remaining cluster-count errors, and \textsc{wrong-assignment} captures valid partitions with incorrect item assignments.

\paragraph{Balanced Hard-pair Selection}
After assigning error categories, we select hard pairs with three goals. 
First, pairs should be difficult for the initial model, as indicated by a small answer-token margin. 
Second, selected pairs should cover different structural error types. 
Third, pairs should be balanced across clustering perspectives so that DPO does not overfit to a small number of frequent perspectives. 
We cap the number of rejected answers per prompt and use perspective- and error-balanced sampling to construct the final DPO dataset.

\paragraph{DPO Objective}
The policy model $\pi_\theta$ and reference model $\pi_{\mathrm{ref}}$ are initialized from the same RD checkpoint. 
The reference model is frozen during DPO. 
For each selected preference pair, we compute answer-token mean log-probabilities:
\begin{equation}
\Delta_\pi
=
\log \pi_\theta(y^+\mid x)
-
\log \pi_\theta(y^-\mid x),
\end{equation}
\begin{equation}
\Delta_{\mathrm{ref}}
=
\log \pi_{\mathrm{ref}}(y^+\mid x)
-
\log \pi_{\mathrm{ref}}(y^-\mid x).
\end{equation}
The scaled DPO logit is
\begin{equation}
z = \beta L_{\mathrm{ref}}(\Delta_\pi-\Delta_{\mathrm{ref}}),
\end{equation}
where $\beta$ controls preference strength and $L_{\mathrm{ref}}$ is the average answer length in the DPO training set. 
The final loss is
\begin{equation}
\mathcal{L}_{\mathrm{DPO}}
=
-\log \sigma(z).
\end{equation}

All log-probabilities are computed only over the canonical answer block. 
This design aligns the model toward better clustering decisions while avoiding direct optimization over long free-form reasoning text.

\section{Baseline Implementation Details}
\label{app:baselines}
We evaluate a set of transcript-embedding clustering baselines: audio-to-text conversion followed by embedding-based clustering. For each benchmark instance in the $\mathrm{\textbf{L}}_0$, $\mathrm{\textbf{L}}_1$, and $\mathrm{\textbf{L}}_2$ splits, the baseline first obtains a transcript for each audio segment. For each corpus, the default entrypoints use \texttt{OpenAI Whisper} to transcribe the audio on the fly; the latter transcripts are normalized by removing speaker-role prefixes and collapsing whitespace. Given the natural-language clustering perspective, each transcript is then converted into a task-conditioned text input using model-specific templates. We run four embedding backbones defined by the official baseline configuration: \texttt{hkunlp/instructor-large}, \texttt{BrandonZYW/llama-2-7b-InBedder}, \texttt{Qwen/Qwen3-Embedding-0.6B}, and \texttt{sentence-transformers/all-MiniLM-L6-v2}. For Instructor, the perspective is provided as an embedding instruction; for Qwen3-Embedding and all-MiniLM, the perspective and transcript are concatenated into a single task-aware query; for InBedder, the transcript and instruction are formatted as an input--instruction prompt and the final hidden representation after short generation is used as the embedding. All embeddings are L2-normalized by default. We then apply two standard clustering algorithms, K-Means and GMM, to the embeddings. Unless otherwise specified, the number of clusters is set to the number of gold clusters, giving these baselines an oracle cluster-count setting. K-Means is run with random seed 43 and automatic initialization when supported by the installed scikit-learn version; GMM is run with diagonal covariance, regularization coefficient $10^{-6}$, and the same random seed. The resulting cluster assignments are compared against the gold labels using ARI and V-measure. Failed embedding or clustering runs are treated as invalid predictions and receive zero score in the aggregate evaluation.

For transcription-based LLM clustering, we follow the same ASR strategy as aforementioned, and use the following prompt to generate a clustering result by taking a collection of transcribed scripts.
\begin{promptbox}[Prompt for transcribed scripts reasoning]
You are a clustering assistant to do audio clustering.
Given a clustering goal and a list of indexed audio:
First, listen to all audio recordings and think how can they be clustered based on the goal, determine the total number of clusters.
Then think about how to assign all audio recordings into these clusters.
Check the answer format before giving the final answer: every item must be assigned to exactly one cluster, and no item should appear in multiple clusters or be missing.
The reasoning and answer must be enclosed within <think> </think> and <answer> </answer> tags, respectively.
Final Output Format should be:
<think> assistant's reasoning process here </think>
<answer>
Total clusters: [N].
cluster1: [item_numbers separated by commas].
cluster2: [item_numbers separated by commas].
...
</answer>
Now, please follow the format for the following clustering task:
Goal: {CLUSTERING_PERSPECTIVE}
The following are transcribed texts from indexed audio segments. Please cluster these segments based on the transcribed content and task goal.
Transcribed Audio Text: {TEXTS}
  \end{promptbox}

For LALM baselines, we use a direct audio prompting protocol that requires the model to perform clustering from native audio inputs rather than from transcripts. The prompt explicitly instructs the model to listen to each audio segment in order, where the $i$-th audio segment corresponds to item $i$ under 1-based indexing. This design ensures that the model conditions its prediction on the original speech signal and can exploit both linguistic content and paralinguistic cues when they are relevant to the specified perspective. To improve output consistency and facilitate automatic parsing, the prompt further requires the model to verify that every item is assigned to exactly one cluster, with no duplicated or missing items. The model is asked to produce its response in a structured format with separate reasoning and answer fields enclosed by \texttt{<think>} and \texttt{<answer>} tags. The final answer must report the inferred number of clusters and list the item indices assigned to each cluster. Unlike embedding-based baselines that assume an oracle number of clusters, this prompting protocol requires the LALM to jointly infer both the cluster count and the cluster assignments from the provided audio collection.

\begin{promptbox}[Prompt for native LALMs]
You are a clustering assistant for native audio inputs.
You will receive a clustering goal and multiple audio segments attached as audio (not as transcripts). Listen to each segment in order; segment index i corresponds to item i (1-based numbering in the final answer).
Do not claim that you only have text--use the provided audio. If audio parts are present in the user message, you must base clustering on listening.
Check the answer format before giving the final answer:
Every item must be assigned to exactly one cluster,
and no item should appear in multiple clusters or be missing.
The reasoning and answer must be enclosed within <think> </think> and <answer> </answer> tags, respectively.
Final Output Format should be:
<think> assistant's reasoning process here </think>
<answer>
Total clusters: [N].
cluster1: [item_numbers separated by commas].
cluster2: [item_numbers separated by commas].
...
</answer>
Now, please follow the format for the following clustering task:
Goal: {CLUSTERING_PERSPECTIVE}
Audio: {SPEECH_CLIPS}
  \end{promptbox}

\section{Implementation Config Details}
\label{app:config}
\subsection{Reasoning Distillation}
\label{app:RD}
For the reasoning distillation, we train on the distilled trainset, which contains 1876 examples. 
The model is initialized from the 7B Audio Flamingo 3 and fine-tuned with LoRA for parameter-efficient adaptation. Training is conducted in \texttt{bfloat16}, with gradient checkpointing enabled to reduce memory usage, together with FlashAttention and DeepSpeed-based memory optimization. The micro-batch size is set to 1, and the run uses a cosine learning-rate schedule with a peak learning rate of $5\times10^{-5}$. We train for 6 epochs. This experiment is designed to preserve the full reasoning supervision signal during distillation and to improve the model's reasoning and response quality on audio-language instruction-following tasks.
\subsection{Direct Preference Optimization}
\label{app:DPO}
We perform scaled-mean DPO optimization on a screened on-policy preference set of 1,075 training pairs constructed from the held-in perspectives. 
Training is run with the official \texttt{trl.DPOTrainer} on 1 node with 4 GPUs, with per-device batch size 1 and gradient accumulation 3, giving an effective global batch size of 12. The policy and reference are both initialized from the reasoning distilled model, and optimization uses scaled-mean DPO with $\beta = 0.5$ and a fixed length scale $L_{\mathrm{ref}} = 34.0$. We train for 500 steps (with an epoch cap of 20.0), using a constant-with-warmup schedule with 5 warmup steps and learning rate $5 \times 10^{-6}$. Training uses \texttt{bfloat16}, gradient checkpointing, max grad norm 1.0, and a LoRA adapter applied to attention modules in the last 16 layers with rank 32, alpha 32, and dropout 0. The objective is computed on answer-only tokens, excluding prompt, padding, and non-answer tokens.

\section{Evaluation Metrics}
\label{app:metrics}
Evaluating clustering quality is a fundamental yet non-trivial problem, as cluster labels are inherently unordered and lack a direct correspondence with ground-truth class labels. Consequently, effective evaluation metrics must be invariant to label permutations and capable of capturing different aspects of clustering structure. Broadly, external clustering metrics can be categorized into \textit{information-theoretic} measures and \textit{pair-counting} measures. In this work, we adopt two widely used and complementary metrics: \textit{V-measure}~\cite{rosenberg2007v}, which is grounded in information theory, and \textit{Adjusted Rand Index (ARI)}~\cite{hubert1985comparing}, which is based on pairwise assignment consistency.

\noindent \textbf{V-measure} is an entropy-based metric designed to quantify the agreement between predicted clusters and ground-truth classes through two desirable properties: \textit{homogeneity} and \textit{completeness}. Homogeneity measures whether each cluster contains only samples from a single class, thus penalizing cluster impurity. Completeness, on the other hand, evaluates whether all samples belonging to a given class are assigned to the same cluster, penalizing class fragmentation across multiple clusters.

These properties are formalized using conditional entropy. Let $C$ denote the set of ground-truth labels and $K$ denote the set of predicted clusters. The homogeneity score is defined as:
\begin{align}
h = 1 - \frac{H(C \mid K)}{H(C)},
\end{align}
which becomes $1$ when each cluster contains only one class (i.e., zero conditional entropy). Similarly, completeness is defined as:
\begin{align}
c = 1 - \frac{H(K \mid C)}{H(K)},
\end{align}
which reaches $1$ when all members of a class are assigned to a single cluster.

To balance these two criteria, V-measure computes their harmonic mean:
\begin{align}
V = (1 + \beta) \cdot \frac{h \cdot c}{\beta \cdot h + c},
\end{align}
where $\beta$ controls the relative importance of completeness over homogeneity (typically $\beta = 1$). The harmonic mean ensures that a high V-measure score is achieved only when both homogeneity and completeness are simultaneously high. The resulting score lies in $[0,1]$, with higher values indicating better alignment between clustering and ground truth.

\noindent \textbf{Adjusted Rand Index} evaluates clustering quality by considering all pairs of samples and measuring how consistently they are assigned in both the predicted clustering and the ground-truth labeling. Specifically, a pair of samples can either be assigned to the same cluster or to different clusters. ARI counts agreements and disagreements between the two partitions over all $\binom{n}{2}$ possible pairs.

The original Rand Index (RI) computes the fraction of agreeing pairs; however, it does not account for agreements that may occur by chance, especially when the number of clusters is large or unbalanced. ARI addresses this limitation by introducing a chance-adjusted normalization. Let $n_{ij}$ denote the number of samples assigned to ground-truth class $i$ and predicted cluster $j$, and define $a_i = \sum_j n_{ij}$ and $b_j = \sum_i n_{ij}$. The ARI is computed as:
\begin{align}
\text{ARI} =
\frac{
\sum_{ij} \binom{n_{ij}}{2}
- \frac{\sum_i \binom{a_i}{2} \sum_j \binom{b_j}{2}}{\binom{n}{2}}
}{
\frac{1}{2} \left(
\sum_i \binom{a_i}{2}
+ \sum_j \binom{b_j}{2}
\right)
- \frac{\sum_i \binom{a_i}{2} \sum_j \binom{b_j}{2}}{\binom{n}{2}}
}.
\end{align}
This normalization ensures that the expected ARI of random clusterings is approximately $0$, providing a meaningful baseline. The ARI ranges from $-1$ to $1$, where $1$ indicates perfect agreement, $0$ corresponds to random assignments, and negative values indicate worse-than-random clustering.

\noindent \textbf{Complementary Perspectives}
V-measure and ARI capture complementary aspects of clustering quality. V-measure emphasizes global information consistency and is particularly sensitive to over-segmentation (low completeness) and mixed clusters (low homogeneity). In contrast, ARI focuses on pairwise consistency and is sensitive to both cluster size distribution and the relative placement of individual samples. Using both metrics provides a more comprehensive evaluation, especially in complex settings such as audio multi-perspective clustering, where both semantic purity and structural consistency are crucial.

\section{More Analysis}
\label{app:more_analysis}

\paragraph{Comparison with ASR-based Pipelines}
Tables~\ref{tab:ARI} and~\ref{tab:V-measure} show a clear gap between \Mname{} and ASR-based pipelines.
Among embedding-based methods, the strongest overall result is achieved by Instructor with K-Means, reaching 17.01 ARI and 52.66 V-measure.
Replacing the embedding-and-clustering stage with an LLM improves performance substantially: Whisper+GPT-4o obtains 28.38 overall ARI and 60.42 overall V-measure.
However, \Mname{} still outperforms this ASR+LLM baseline by +16.39 ARI points and +13.01 V-measure points.
This suggests that simply transcribing speech into text is insufficient for audio multi-perspective clustering.
By directly operating on audio inputs, \Mname{} can exploit acoustic and paralinguistic cues that may be weakened or discarded during ASR.

\paragraph{Comparison with Native LALMs}
\Mname{} also substantially outperforms off-the-shelf native audio-language models.
Among these baselines, GPT-audio-1.5 is the strongest overall system, achieving 31.78 ARI and 61.81 V-measure.
In contrast, \Mname{} reaches 44.77 ARI and 73.43 V-measure, improving over GPT-audio-1.5 by +12.99 ARI points and +11.62 V-measure points.
\Mname{} obtains the best ARI across all 12 corpus-level evaluation settings and the best V-measure on 11 out of 12 settings.
These results indicate that general-purpose audio-language models do not automatically acquire reliable set-partitioning behavior, and that task-specific post-training is important for perspective-conditioned clustering.

\paragraph{Failure Mode of Audio Flamingo 3}
Audio Flamingo 3 obtains very low scores in our evaluation, which is mainly due to its inability to reliably follow the required clustering-output protocol. 
In many cases, the model produces free-form responses that cannot be parsed into a valid partition, such as missing items, duplicated assignments, or outputs without an explicit cluster structure. 
To avoid underestimating the model solely due to rigid formatting constraints, we additionally applied an LLM-based extractor to recover clustering assignments from its responses when possible. 
However, the recovered partitions still led to very low clustering scores, suggesting that the failure is not merely a parsing artifact. 
Rather, off-the-shelf Audio Flamingo 3 lacks the task-specific behavior needed for perspective-conditioned set partitioning: it must interpret the clustering perspective, compare multiple audio segments jointly, infer the number of clusters, and produce a complete valid partition. 
This observation further motivates our reasoning distillation and preference optimization stages, which explicitly train \Mname{} to generate valid and clustering-aligned outputs.

\paragraph{Corpus-level Observations.}
The gains of \Mname{} are consistent across all four corpora, but the nature of the improvement differs by domain.
On ECHR and S\&P 500, \Mname{} substantially improves both ARI and V-measure, suggesting stronger ability to organize long-form legal and financial speech under different clustering perspectives.
On Banking77, \Mname{} achieves especially strong ARI improvements, indicating that the model can distinguish fine-grained intent-oriented spoken utterances while also using audio-side information when required.
On MultiWOZ, \Mname{} also outperforms all baselines in ARI and achieves the best V-measure across all three evaluation levels.
This suggests that the proposed training pipeline improves not only paralinguistic perception but also dialogue-domain clustering, where the model must reason over interaction structure and pragmatic intent.

We also train the model with using Qwen2.5-omni as the base model.
The evaluation results are included in Tab.~\ref{tab:task_group_sft_dpo}.

\begin{table*}[t]
\centering
\small
\setlength{\tabcolsep}{2.5pt}
\renewcommand{\arraystretch}{1.08}
\resizebox{\textwidth}{!}{
\begin{tabular}{ll*{25}{c}}
\toprule
\toprule
\multirow{2}{*}{Metric} &
\multirow{2}{*}{Training} &
\multicolumn{5}{c}{ECHR} &
\multicolumn{5}{c}{S\&P 500} &
\multicolumn{5}{c}{Banking77} &
\multicolumn{5}{c}{MultiWOZ} &
\multicolumn{5}{c}{Overall} \\
\cmidrule(lr){3-7}
\cmidrule(lr){8-12}
\cmidrule(lr){13-17}
\cmidrule(lr){18-22}
\cmidrule(lr){23-27}
& &
BG & Emo. & Spk. & Gen. & Reason &
BG & Emo. & Spk. & Gen. & Reason &
BG & Emo. & Spk. & Gen. & Reason &
BG & Emo. & Spk. & Gen. & Reason &
BG & Emo. & Spk. & Gen. & Reason \\
\midrule

\multirow{2}{*}{ARI}
& \cellcolor{uciblue!5} RD SFT
& \cellcolor{uciblue!5} 0.2947 & \cellcolor{uciblue!5} 0.2809 & \cellcolor{uciblue!5} 0.2682 & \cellcolor{uciblue!5} 0.2806 & \cellcolor{uciblue!5} 0.3872
& \cellcolor{uciblue!5} 0.3754 & \cellcolor{uciblue!5} 0.3096 & \cellcolor{uciblue!5} 0.3212 & \cellcolor{uciblue!5} 0.3539 & \cellcolor{uciblue!5} 0.3968
& \cellcolor{uciblue!5} 0.3380 & \cellcolor{uciblue!5} 0.2616 & \cellcolor{uciblue!5} 0.3025 & \cellcolor{uciblue!5} 0.3803 & \cellcolor{uciblue!5} 0.4749
& \cellcolor{uciblue!5} 0.3127 & \cellcolor{uciblue!5} 0.3664 & \cellcolor{uciblue!5} 0.0000 & \cellcolor{uciblue!5} 0.3823 & \cellcolor{uciblue!5} 0.4067
& \cellcolor{uciblue!5} 0.3302 & \cellcolor{uciblue!5} 0.3046 & \cellcolor{uciblue!5} 0.2230 & \cellcolor{uciblue!5} 0.3493 & \cellcolor{uciblue!5} 0.4164 \\

& RD+DPO
& 0.3163 & 0.2882 & 0.2727 & 0.2993 & 0.3821
& 0.3965 & 0.3169 & 0.3293 & 0.3554 & 0.4014
& 0.3818 & 0.2437 & 0.3448 & 0.3786 & 0.4507
& 0.3407 & 0.3557 & 0.0000 & 0.3724 & 0.4010
& 0.3588 & 0.3012 & 0.2367 & 0.3514 & 0.4088 \\

\midrule

\multirow{2}{*}{V-measure}
& \cellcolor{uciblue!5} RD SFT
& \cellcolor{uciblue!5} 0.6135 & \cellcolor{uciblue!5} 0.6772 & \cellcolor{uciblue!5} 0.6212 & \cellcolor{uciblue!5} 0.5756 & \cellcolor{uciblue!5} 0.8025
& \cellcolor{uciblue!5} 0.6460 & \cellcolor{uciblue!5} 0.6879 & \cellcolor{uciblue!5} 0.6294 & \cellcolor{uciblue!5} 0.5866 & \cellcolor{uciblue!5} 0.7717
& \cellcolor{uciblue!5} 0.5475 & \cellcolor{uciblue!5} 0.6769 & \cellcolor{uciblue!5} 0.4725 & \cellcolor{uciblue!5} 0.5801 & \cellcolor{uciblue!5} 0.8592
& \cellcolor{uciblue!5} 0.5115 & \cellcolor{uciblue!5} 0.7023 & \cellcolor{uciblue!5} 0.0000 & \cellcolor{uciblue!5} 0.5134 & \cellcolor{uciblue!5} 0.6240
& \cellcolor{uciblue!5} 0.5796 & \cellcolor{uciblue!5} 0.6861 & \cellcolor{uciblue!5} 0.4308 & \cellcolor{uciblue!5} 0.5639 & \cellcolor{uciblue!5} 0.7643 \\

& RD+DPO
& 0.6221 & 0.6827 & 0.6223 & 0.5864 & 0.8000
& 0.6595 & 0.6920 & 0.6329 & 0.5837 & 0.7689
& 0.5834 & 0.6660 & 0.4921 & 0.5779 & 0.8485
& 0.5384 & 0.6985 & 0.0000 & 0.5141 & 0.6238
& 0.6008 & 0.6848 & 0.4368 & 0.5655 & 0.7603 \\

\bottomrule
\bottomrule
\end{tabular}
}
\caption{Performance of using Qwen2.5-omni as the base model. BG, Emo., Spk., Gen., and Reason denote background noise, emotion, speaker count, gender, and the average over linguistic reasoning perspectives.}
\label{tab:task_group_sft_dpo}
\end{table*}

\subsection{Case Study}
\label{sec:case-study}
We present a case study from the ECHR domain to illustrate perspective-conditioned audio clustering. The model is asked to cluster legal cases by the main relationship between the applicant and the responsible actor(s), especially distinguishing direct State conduct from protection, enforcement, or regulatory failures. Although transcripts are shown for readability, \Mname{} receives the original audio recordings during inference.

As shown in Fig.~\ref{fig:case}, \Mname{} separates the seven cases into four relation-based groups: administrative or regulatory disputes, direct coercive State conduct, a private dispute involving judicial protection, and detention-condition complaints. This grouping reflects the intended perspective rather than superficial topical similarity. For example, police abuse and military police shooting are grouped together as direct State force, while customs seizure, land-transfer approval, and contaminated blood-product compensation are grouped as administrative or regulatory responsibility.

This case study suggests that \Mname{} can infer abstract relational structures from native audio inputs, supporting flexible clustering under user-specified perspectives.

\begin{figure*}[h!]
  \centering
  \begin{tcolorbox}[
      colback=uciblue!5,
      colframe=black,
      boxrule=0.3mm,
      arc=3mm,
      width=\textwidth
    ]
\small
\raggedright

You are a clustering assistant for native audio inputs.
You will receive a clustering goal and multiple audio segments attached as audio (not as transcripts). Listen to each segment in order; segment index i corresponds to item i (1-based numbering in the final answer).
Do not claim that you only have text--use the provided audio. If audio parts are present in the user message, you must base clustering on listening.
\medskip

Check the answer format before giving the final answer:
Every item must be assigned to exactly one cluster, and no item should appear in multiple clusters or be missing.
The reasoning and answer must be enclosed within \texttt{<think>} \texttt{</think>} and \texttt{<answer>} \texttt{</answer>} tags, respectively.
\medskip

Final Output Format should be:
\texttt{<think>} assistant's reasoning process here \texttt{</think>}
\texttt{<answer>}
Total clusters: [N].\medskip

Cluster1: [item\_numbers separated by commas].\medskip

Cluster2: [item\_numbers separated by commas].\medskip

...
\texttt{</answer>}
\medskip

Now, follow the format for the following clustering task:
Goal: Cluster these legal cases by the main relationship between applicant and responsible actor(s), distinguishing direct State conduct from protection/enforcement failures.
Audio: \texttt{SPEECH\_1}, \texttt{SPEECH\_2}, \texttt{SPEECH\_3}, \texttt{SPEECH\_4}, \texttt{SPEECH\_5}, \texttt{SPEECH\_6}, \texttt{SPEECH\_7}
\medskip

----------------------------------------------------------------------------------------------------------------------------------------------
\medskip

\texttt{<think>} I’m mainly hearing two big relationship types: cases where the harm comes straight from State agents’ actions, versus cases where the State is involved more as a regulator/court system (or as custodian) rather than as the immediate perpetrator. Items 1 and 5 clearly sit together: both are on-the-spot encounters with police/military police using force. In 1 it’s beating in custody plus delayed medical help; in 5 it’s live fire during an arrest attempt leading to deaths. These feel like direct operational conduct by security forces, not just a regulatory decision. Item 2 is also about State control, but it’s not a single arrest/force incident—it’s ongoing detention conditions complained of against the detaining authority. That separates it from 1/5 into a custody-conditions bucket. Then there’s a set where the applicant is essentially fighting the State as administrator/regulator: 7 (customs seizure/damage and compensation litigation), 3 (land transfer approval blocked by property-transaction authorities), 4 (infected blood products and administrative liability/compensation). I hesitated about 7 because a private company appears, but the core dispute is still with customs and State responsibility. Finally, 6 is different: a private insurance dispute where the State’s role is providing a court process (with delay issues), not being the opposing actor. \texttt{</think>} \texttt{<answer>} 
Total clusters: [4].\medskip

Cluster 1: [3, 4, 7]\medskip

Cluster 2: [1, 5]\medskip

Cluster 3: [6]\medskip

Cluster 4: [2]\medskip
\texttt{</answer>}
  \end{tcolorbox}
  \vspace{-1em}
  \caption{Case study}
  \vspace{-1em}
  \label{fig:case}
\end{figure*}

\texttt{SPEECH\_1}: [\textbf{Serena}: \textit{In May 2005, the applicant was stopped on the street by police officers and taken into custody at the Sovetskiy district police station in Orsk. He tried to escape but was assaulted by the officers, who kicked him in the stomach.}
\textbf{Vivian}: \textit{That assault caused blunt abdominal trauma, including a ruptured intestine and serious health damage. He lost consciousness and was placed in a cell, where the police ignored his requests for medical help.}
\textbf{Serena}: \textit{The next day, he was finally hospitalized with internal bleeding and spent six weeks receiving care. Forensic reports confirmed his injuries were caused by blunt force trauma, ruling out any accidental causes.}
\textbf{Vivian}: \textit{Following this, the applicant brought a civil claim against the State authorities for ill-treatment and an ineffective investigation. The Leninskiy District Court partially granted compensation, finding the injuries occurred in police custody with no alternative explanation provided.}
\textbf{Serena}: \textit{That judgment was upheld on appeal, highlighting issues like direct police encounters, use of force by agents, physical abuse, and deprivation of liberty.}
\textbf{Vivian}: \textit{Yes, and the case also emphasized the failure of authorities to properly investigate the ill-treatment, which led to the civil litigation and compensation awarded to the applicant.}
]

\noindent\texttt{SPEECH\_2}: [\textbf{Sohee}: \textit{The applicants, who are currently deprived of their liberty, have filed complaints against the detaining authority.}
\textbf{Serena}: \textit{Yes, their main concern is the inadequate and inhuman conditions they face while in detention.}
\textbf{Sohee}: \textit{Exactly. The details about each applicant and their specific applications are outlined in the appended table.}
\textbf{Serena}: \textit{Their grievances focus especially on the custody conditions and the care they receive, which they describe as poor.}
\textbf{Sohee}: \textit{They also highlight the ill-treatment they have suffered, which stems from the harsh detention environment.}
\textbf{Serena}: \textit{So, overall, their complaints paint a troubling picture of neglect and mistreatment within the detention facilities.}]

\noindent\texttt{SPEECH\_3}: [\textbf{Sohee}: \textit{The applicant, a U.S. citizen living in Munich, wanted to build a holiday home in Hopfgarten, Austria. She began negotiating to buy land there back in 1971.}
\textbf{Dylan}: \textit{Her purchase contract needed approval under the Tyrolean Real Property Transactions Act. At first, the local authority gave the green light for the sale.}
\textbf{Ono\_Anna}: \textit{But then the Real Property Transactions Officer challenged this decision by appealing to the Regional Authority. The concern was that with 110 foreign landowners already in Hopfgarten, this sale might lead to foreign domination.}
\textbf{Vivian}: \textit{The Regional Authority agreed and refused to approve the transfer. They argued that the purchase could harm social and economic interests and noted the land was intended for a holiday home, not farming.}
\textbf{Sohee}: \textit{The applicant then took her case to the Constitutional Court, claiming her property rights and right to a fair court were violated. She also argued the Regional Authority wasn’t independent.}
\textbf{Dylan}: \textit{However, the Constitutional Court dismissed her appeal. They confirmed the Regional Authority’s independence and upheld their decision to block the land transfer.}
\textbf{Ono\_Anna}: \textit{Meanwhile, the applicant and her family lived in Germany with temporary residence permits. She was even willing to apply for Austrian nationality to resolve the issue.}]

\noindent\texttt{SPEECH\_4}: [\textbf{Ono\_Anna}: \textit{Mr. Jean-Marc Pailot, born in 1952, is a French clerical worker and haemophiliac who received multiple blood transfusions. On August 27, 1985, he tested positive for HIV. Seeking compensation, he approached the Minister for Solidarity, Health and Social Protection, but his claim was rejected. Mr. Pailot then took his case to the Châlons-sur-Marne Administrative Court, which referred the matter to the Conseil d’Etat. Initially, the Administrative Court held the State liable for infections from non-heat-treated blood products between March 12 and October 1, 1985, but expert evidence could not pinpoint the exact infection date. Appeals followed, involving the Deputy Minister for Health and the Minister of Employment and Social Affairs. Ultimately, the Conseil d’Etat overturned earlier rulings and found the State liable for infections occurring between November 22, 1984, and October 20, 1985, ordering compensation. The case was a legal dispute between an individual and regulatory authorities, with no indication of vulnerability beyond Mr. Pailot’s medical condition, proceeding through administrative judicial review.}]

\noindent\texttt{SPEECH\_5}: [\textbf{Serena}: \textit{On July 19, 1996, two unarmed conscripts, Mr. Angelov and Mr. Petkov, fled detention and were chased by four military police officers sent to arrest them.}
\textbf{Uncle\_Fu}: \textit{Right, and these officers were told to use whatever means necessary. They found the men at their grandmother's house in Lesura, where the fugitives tried to escape through a window and over fences.}
\textbf{Serena}: \textit{Sergeant N. shouted ``Stop, military police!'' but didn’t fire any shots. Then Major G. gave warnings and fired multiple shots with his automatic rifle, aiming at their feet to stop them.}
\textbf{Uncle\_Fu}: \textit{Witnesses confirmed Major G. was the only one who fired live rounds. The others fired shots into the air. Both men were wounded but alive when taken to the hospital, though they died on the way.}
\textbf{Serena}: \textit{This incident involved direct police use of force that resulted in death, leading to complaints about police conduct. The officers knew who the fugitives were, and the shooting happened during a direct encounter.}]

\noindent\texttt{SPEECH\_6}: [\textbf{Ono\_Anna}: \textit{The applicant was injured in a work accident back in 1993 and decided to sue the insurance company ZT for damages in civil court.}
\textbf{Sohee}: \textit{Between 1996 and 1999, she submitted eight preliminary filings, presented evidence, and requested six times that a hearing date be set. Eventually, three hearings took place, none adjourned at her request, and a medical expert was appointed to assess the case.}
\textbf{Eric}: \textit{The initial judgment partially upheld her claim, but in 1997, the presiding judge was replaced. After ZT appealed, the higher court partly allowed the appeal and sent the case back for re-examination.}
\textbf{Ono\_Anna}: \textit{Following that, the applicant filed four rush notices to speed up the process. There were also decisions on costs, and ZT's appeals against those were rejected by 2002.}
\textbf{Sohee}: \textit{Throughout the dispute, which involved private parties, the court provided protection. Although there were procedural delays and remands, no specific vulnerability of the applicant was indicated during the proceedings.}]

\noindent\texttt{SPEECH\_7}: [\textbf{Uncle\_Fu}: \textit{In 1997, the applicant was detained on suspicion of drug trafficking, and customs officers seized his car, belongings, documents, and money but refused to make an inventory. The damaged car was transferred by the Customs Service to a private company, which returned it missing parts, with torn documents and lost money. The applicant sued the Sverdlovsk Customs Service for compensation for both financial and non-financial damages, starting civil proceedings that were suspended while criminal investigations against two customs officers allegedly responsible for the damage were ongoing. The courts initially allowed some claims but later rejected them, with the criminal case still unresolved. After the applicant died in 2007, his mother and daughter joined the civil case, represented by his sister, and hearings resumed in 2009. This dispute involves an individual challenging a regulatory authority over property loss and inadequate compensation, with the civil litigation still ongoing in the first instance court.}]

\end{document}